*Article*

# Using Density Functional Theory to Model Realistic TiO₂ Nanoparticles, Their Photoactivation and Interaction with Water

**Daniele Selli, Gianluca Fazio and Cristiana Di Valentin ***

Dipartimento di Scienza dei Materiali, Università di Milano-Bicocca, via R. Cozzi 55, 20125 Milano, Italy;
daniele.selli@unimib.it (D.S.); g.fazio3@campus.unimib.it (G.F.)
* Correspondence: cristiana.divalentin@unimib.it

**Abstract:** Computational modeling of titanium dioxide nanoparticles of realistic size is extremely relevant for the direct comparison with experiments but it is also a rather demanding task. We have recently worked on a multistep/scale procedure to obtain global optimized minimum structures for chemically stable spherical titania nanoparticles of increasing size, with diameter from 1.5 nm (~300 atoms) to 4.4 nm (~4000 atoms). We use first self-consistent-charge density functional tight-binding (SCC-DFTB) methodology to perform thermal annealing simulations to obtain globally optimized structures and then hybrid density functional theory (DFT) to refine them and to achieve high accuracy in the description of structural and electronic properties. This allows also to assess SCC-DFTB performance in comparison with DFT(B3LYP) results. As a further step, we investigate photoexcitation and photoemission processes involving electron/hole pair formation, separation, trapping and recombination in the nanosphere of medium size by hybrid DFT. Finally, we show how a recently defined new set of parameters for SCC-DFTB allows for a proper description of titania/water multilayers interface, which paves the way for modeling large realistic nanoparticles in aqueous environment.

## 1. Introduction

TiO₂ nanoparticles are fundamental building blocks in photocatalysis [1–4]. Their theoretical description is indeed relevant and requires the size of the model to be as realistic as possible, for direct comparison with experimental samples.

TiO₂ nanoparticles are most typically obtained from sol-gel synthesis. Several studies have proven that shape and size can be successfully tailored by controlling the conditions of preparation and by using ad-hoc surface chemistry [5–7]. The minimum energy shape was predicted by Barnard et al. [8] by Wulff construction, for dimensions below 10 nm, to be a decahedron in the anatase phase, exposing mainly (101) and small (001) facets. However, growth determining factors are pH and particle density. An excessive dilution may cause a partial dissolution of titania nanocrystals leading to the formation of spherical nanoparticles [9]. Those, analogously to nanotubes and nanorods, are characterized by a high curvature profile and, thus, expected to be more reactive towards molecular adsorption.

The majority of the computational first-principles studies are devoted to bulk or surface slabs of anatase TiO₂ [10]. Few works have dealt with the decahedral faceted nanoparticles [11–16] but none with spherical ones. Modelling nanoparticles of realistic size (few nanometers) by first-principles calculations is very demanding and a global optimization is hardly feasible [17].

Ours is a multistep/scale approach [18] where we propose first to apply a less expensive but still rather accurate method based on density functional theory (DFT), which is the self-consistent-charge density functional tight-binding (SCC-DFTB) [19], to perform a global structure optimization search of the nanoparticles; then to run a further DFT relaxation to determine structural and electronic properties with first-principles level accuracy. For the latter, we use hybrid functionals since they are known to better describe electronic structure details of $TiO_2$ materials [20–22]. SCC-DFTB has been demonstrated to be a powerful tool for the quantum mechanics study of many system involving $TiO_2$ [23–26]. The method retains most of the physics of standard DFT at an extremely reduced computational cost.

Furthermore, we would like to describe the interaction of such nanoparticles with light and their photoactivation producing energy carriers (excitons) and charge carriers (electrons and holes). The aim is to improve the general understanding of the processes at the basis of light energy conversion into chemical species with intrinsic redox potential that are those triggering the redox reaction at the oxide surface [14,27–30].

It is generally accepted that water, as the surrounding environment where titania nanoparticle work in photocatalytic processes, plays an active role [31–39]. It is, therefore, fundamental to describe accurately the dynamical water layers arrangement on the surface and how water molecules may enter the photoactivated reaction chain [40].

In the following, we will present a critical review of our work, relative to the topics highlighted and discussed above: in Section 2, we present the Computational methodology; in Section 3, we describe how to obtain realistic spherical nanoparticles models; in Section 4, we discuss the description of the photoexcitation processes; and, in Section 5, we analyze how the water environment can be modeled with sufficient accuracy.

## 2. Computational Details

To tackle the surface complexity of the $TiO_2$ spherical nanoparticles and maintain a high degree of accuracy, different levels of theory are necessary. Nowadays, density functional theory (DFT) is the most used method to properly describe equilibrium geometries and electronic structures. However, many interesting features of the system are accessible only at certain size and time scale. For example, to explore the potential energy surface related to the different configurations of the $TiO_2$ spherical nanoparticles through molecular dynamics and simulated annealing processes, an approximated method has to be used. The self-consistent-charge density functional tight-binding (SCC-DFTB) approach is a DFT-based quantum mechanical method, which retains a quantum description of the system at a considerably reduced computational cost. Thus, in addition to geometry optimizations and electronic structure calculations, SCC-DFTB also enables molecular dynamic simulations for large systems with a reasonable time length.

### 2.1. Electronic Structure Calculations

The choice of a specific density functional is based on the aim of the study. Standard generalized-gradient approximation (GGA) functionals may be sufficient to describe equilibrium geometries or adsorption energies, however a correct description of the electronic structure of semiconducting oxides requires the inclusion of a certain portion of exact exchange. The use of such hybrid functionals in a plane-wave code is extremely cumbersome, thus localized basis function codes are preferred. The CRYSTAL14 code [41] has been used for most of the density functional theory (DFT) calculations, employing all-electron Gaussian basis sets [O 8-411(d1) Ti 86-411 (d41) and H 511(p1)] and the B3LYP [42,43] and the HSE06 [44] hybrid functionals. For periodic systems, reference DFT calculations were carried out with the Quantum ESPRESSO [45] simulation package, using the PBE functional [46], ultrasoft Vanderbilt pseudopotentials and a plane-wave basis set with a cut off of 30 Ry (300 Ry for the charge density).

The optimized lattice parameters for bulk $TiO_2$ anatase are 3.789 Å and 3.766 Å for $a$ and 9.777 Å and 9.663 Å for $c$, respectively, for B3LYP and HSE06, which are in good agreement with the experimental values [47].

A $6\sqrt{2} \times 6\sqrt{2} \times 1$ $TiO_2$ anatase bulk supercell with 864 atoms was employed to model the exciton and the related distortions in the bulk. The (101) anatase surface has been taken as a reference for: (i) surface energies; and (ii) interaction with water. (i) We employed the CRYSTAL14 code and a minimal cell slab of ten triatomic layers with 60 atoms, where the periodicity was set along the [$10\bar{1}$] and [010] directions and not in the direction perpendicular to the surface. (ii) A $1 \times 2$ supercell three-triatomic-layer slab (72 atoms) has been used within the Quantum ESPRESSO code, where the replicas in the direction perpendicular to the surface were separated by 20 Å in order to avoid any interaction between images. For the k-point sampling, a $1 \times 1 \times 6$, a $8 \times 8 \times 1$ and a $2 \times 2 \times 1$ Monkhorst–Pack grid was used for the bulk, the minimal slab cell and the $1 \times 2$ slab supercell, respectively.

Anatase $TiO_2$ nanospheres have been carved from a bulk supercell following the procedure already described in a previous work by some of us [17]. Nanoparticles have been considered as molecules in the vacuum with no periodic boundary conditions. Therefore, when an excess electron or hole is introduced in the system, no background of charge is necessary. In the case of open-shell systems, spin polarization is taken into account

Trapping energies ($\Delta E_{trap}$) for excitons, extra electrons and holes are calculated as the total energy difference between the trap optimized geometry and the delocalized solution in the neutral ground state geometry.

The total densities of states (DOS) of the nanoparticles have been simulated with the convolution of Gaussian functions ($\sigma = 0.005$ eV) peaked at the value of the Kohn-Sham energies of each orbital. Projected densities of states (PDOS) are built using the following procedure, based on the molecular orbitals coefficients in the linear combination of atomic orbitals (LCAO): summing the squares of the coefficients of all the atomic orbitals centered on a certain atom type results, after normalization, in the relative contribution of each atom type to a specific eigenstate. Then, the various projections are obtained from the convolution of Gaussian peaks with heights that are proportional to the relative contribution. The zero energy for all the DOS is set to the vacuum level, i.e., the energy of an electron at an infinite distance from the surface of the system.

## 2.2. SCC-DFTB Approach

The self-consistent-charge density functional tight-binding method (SCC-DFTB) is based on the approximation of the Kohn-Sham (KS) DFT formalism. Assuming a second-order expansion of the KS-DFT total energy with respect to the electron density fluctuations, the SCC-DFTB total energy is defined as:

$$E_{tot}^{SCC-DFTB} = \sum_i n_i \varepsilon_i + \frac{1}{2}\sum_{\alpha\beta} v_{rep}^{\alpha\beta}(R_{\alpha\beta}) + \frac{1}{2}\sum_{\alpha\beta} \gamma_{\alpha\beta}\Delta q_\alpha \Delta q_\beta \tag{1}$$

where the first term contains the one-electron energies $\varepsilon_i$ from the diagonalization of an approximated Hamiltonian matrix and represents the attractive part of the energy, whereas the second term approximates the short-range repulsive energy, given by the sum of the pairwise distance-dependent potential $v_{rep}^{\alpha\beta}(R_{\alpha\beta})$ between the pair of atoms $\alpha$ and $\beta$, and $\Delta q_\alpha$ and $\Delta q_\beta$ are the charges induced on the atoms $\alpha$ and $\beta$, which interact through a Coulombic-like potential $\gamma_{\alpha\beta}$. For more information on the details of the SCC-DFTB method, see Refs. [19,48,49].

For all the SCC-DFTB calculations, we employed the DFTB+ simulation package [50]. We initially made used of the "matsci-0-3" Slater–Koster parameters, which have been shown to be well-suited for the study of anatase $TiO_2$ in Ref. [24]. Subsequently, to better describe the titania/water/water interface, we combined the "matsci-0-3" parameters for Ti-O and Ti-Ti interactions with the parameters in the "mio-1-1" set [19] for O-O, O-H and H-H interactions in what we have named as "MATORG" set. Furthermore, we modified the $\gamma_{\alpha\beta}$ function to improve the description of H-bonding, using a hydrogen bonding damping function (HBD), in which a $\zeta = 4$ parameter has been used [51]. In this work, we refer to this HBD modified Slater–Koster parameters set as "MATORG+HBD" [40]. From now on, DFTB will be used as a shorthand for SCC-DFTB.

To perform the simulated annealing procedure, we carried out Born–Oppenheimer DFTB molecular dynamics (MD) within the canonical ensemble (NVT). The integration of the Newton's equations of motion has been done with the Velocity Verlet algorithm, using a relative small time step of 0.5 fs to ensure reversibility. A Nosé–Hoover chain thermostat with a time constant of 0.03 ps was applied to reach the desired temperature during the temperature-annealing simulations. The simulation time length of the annealing processes was made commensurate to the size of the nanosphere. Thus, we used a simulation time up to 45 ps for the 1.5 nm, 24 ps for the 2.2 nm, 14 ps for the 3.0 nm and 11 ps for the 4.4 nm nanosphere. In the case of the titania/water interface, each MD simulation has been performed for 25 ps.

For the molecular dynamics of the titania/water interface, a 1 × 3 supercell anatase (101) slab (108 atoms) with a monolayer (ML), a bilayer (BL) and a trilayer (TL) of water, composed of 6, 12 and 18 water molecules, respectively, was used. The desired temperature of the thermostat was set to a constant low value (160 K) to avoid the desorption of surface water molecules.

### 2.3. Structural Analysis

The extended X-ray adsorption fine structure (EXAFS) simulated spectra has been simulated via a Gaussian convolution of peaks ($\sigma = 0.0005$ Å) centered at the length of the distance between each Ti atom and other atoms (O or Ti) in the first, second, and third coordination shells. Projections have been computed considering only the distances centered on specific Ti atoms with a certain coordination sphere. In the text, we also report the surface-to-bulk ratio, defined as the ratio between the number of Ti and O atoms at the surface of the nanosphere and the number of Ti and O atoms in the bulk.

### 3. Modelling Realistic TiO$_2$ Nanoparticles

We carved TiO$_2$ spherical nanoparticles from large bulk anatase supercells. The radius of the sphere is set to a desired value and only atoms within that sphere are considered, whereas those outside the sphere are removed. Some very low coordinated Ti sites are found to be left at the surface of the model that must be removed or saturated with OH groups; analogously monocoordinated O must be removed or saturated with H atoms. Therefore, we use a number of water molecules to achieve the chemical stability of the nanoparticle. We try to keep the number of water molecules as low as possible. Since we aim at modelling nanoparticles of realistic size, we range from spheres with diameter of 1.5 nm up to 4.4 nm. These contain from 300 up to almost 4000 atoms. The exact stoichiometry of the prepared nanoparticles [(TiO$_2$)$_{101}$·6H$_2$O, (TiO$_2$)$_{223}$·10H$_2$O, (TiO$_2$)$_{399}$·12H$_2$O, and (TiO$_2$)$_{1265}$·26H$_2$O] is reported in Figure 1.

Structural relaxation by geometry optimization from the "as-carved" and chemically stabilized models is not an efficient approach because we found that it leads to local minimum structures, which are far from the global minimum one. For this reason, we have drastically changed our approach and decided to use a less computationally expensive, but still rather accurate, DFT-based method (DFTB) and to run some molecular dynamics simulations starting from the "as-carved" structures at increasing temperature (up to 700 K in some cases). This approach allows moving from the local minimum structure basin, close to the "as-carved" structure, and to further sample the configuration space. The thermally equilibrated structures obtained with this approach, and then fully relaxed, are much more stable to any surface modification (i.e., addition of a molecular adsorbate) because those are true global minima on the potential energy surface of the TiO$_2$ nanospheres. This multi-step procedure can be rather easily and reasonably applied to nanospheres of increasing size, and certainly up to the one with a 4.4 nm diameter (~4000 atoms). Once the fully relaxed nanospheres are prepared, we can investigate structural and electronic properties at the DFTB level of theory. However, to assess the accuracy of DFTB in this specific context, we must perform a benchmark study against hybrid DFT model calculations. Those were obtained by full atomic relaxation starting from the DFTB thermally annealed and optimized spheres. Note that we performed this further DFT(B3LYP) optimization for all four nanospheres. For the very large one (~4000 atoms) it was an extremely expensive procedure, but we consider it worth because a successful benchmark against DFT will assess DFTB reliability for the investigation of structural and electronic properties of spherical $TiO_2$ nanoparticles, as the basis for future developments.

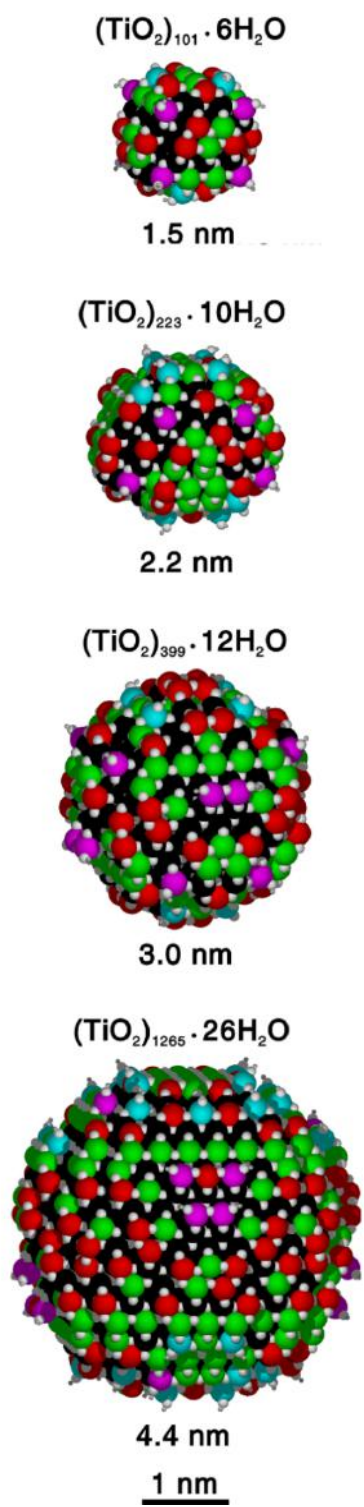

**Figure 1.** DFT(B3LYP) optimized structures, after simulated annealing at DFTB level, of the different nanospheres considered in this work. For each one, the stoichiometry, the approximate diameter and the position of the Ti atoms with different coordination are reported ($Ti_{4c}$ in red, $Ti_{5c}$ in green, $Ti_{6c}$ in black, $Ti_{3c}$(OH) in magenta and $Ti_{4c}$(OH) in cyan).

## 3.1. Structural Properties

Our study is not only meant to validate the DFTB methodology with respect to an accurate hybrid DFT method, but also to highlight how the enhanced curvature present in small $TiO_2$ nanospheres, in analogy to that in nanotubes and nanorods, affects both the coordination of the surface atoms and their geometrical environment. The different coordination types are illustrated in Figure 1 and their relative percentage is reported in Table 1. Flat anatase surfaces and faceted nanoparticles are dominated by $Ti_{6c}$ and $Ti_{5c}$, whereas it is evident that on spherical nanoparticles different types of coordination exist that are expected to play a key role in the processes of chemical adsorption. The structural distortions induced by the nanosize and by the high curvature are investigated through the analysis of the simulated direct space X-ray absorption fine structure spectra (EXAFS) for bulk and for the nanospheres models of different size, as obtained with both DFT(B3LYP) and DFTB calculations (Figure 2). Those provide the distribution of the distances for each Ti atom with the neighboring O or next neighboring Ti atoms. Figure 2a,b show the EXAFS spectra for the bulk case. Here, no distribution is observed since crystalline bulk is characterized by two Ti-O distances (two lines on the left side), which are attributed to equatorial and axial O atoms resulting from a $D_{2d}$ point symmetry at each Ti center [DFT(B3LYP): Ti-$O_{eq}$ = 1.946 Å and Ti-$O_{ax}$ = 2.000 Å; DFTB: Ti-$O_{eq}$ = 1.955 Å and Ti-$O_{ax}$ = 1.995 Å], and single Ti···Ti distances at regular lattice positions [DFT(B3LYP): first shell 3.092 Å, second shell 3.789 Å; DFTB: first shell 3.090 Å, second shell 3.809 Å].

Figure 2c–j reports the EXAFS simulated spectra for the nanospheres of increasing size. With respect to bulk, we register a variety of distances due to the lattice distortion and diversity of coordination sites. For this reason and to improve the level of information provided, we present a convolution of peaks and project it on each type of coordination site. In the case of DFT(B3LYP) (left colomn), the first broad peak on the left, related to first neighbor Ti-O distances, is predominantly made up of $Ti_{6c}$–O bond lenghts. Low coordinated sites contribute to the shorter Ti-O distances than in the bulk (red, yellow, brown lines), whereas $Ti_{5c}$ and $Ti_{6c}$ species contribute to the range of bulk values and to the longer Ti-O distances up to 2.2–2.3 Å (blue, green lines). It is evident that, as the size increases, the relative portion of $Ti_{6c}$ sites increases (blue line) with a distribution of Ti-O distances that becomes increasingly more peaked at the bulk values (see Figure 2i). It is clear that the EXAFS spectrum of the nanosphere with a diameter of 4.4 nm already quite largely resembles that of bulk anatase, since the surface-to-bulk ratio (0.43) is rather reduced with respect to the other nanospheres (1.70 > 0.94 > 0.83). The other peaks, for the second and third coordination spheres of Ti···Ti and of Ti···O, are also quite broad, except for the largest nanosphere, where they are rather sharp and centered at the bulk values. In the case of DFTB (Figure 2b, d, f, h, j), similar considerations hold, although a noticeable difference is that the larger Ti-$O_{ax}$ bonds concentrate at the value of about 2.25 Å. This is a surface distortion effect, which becomes progressively lower as the size of the nanosphere increases.

**Table 1.** Amount of Ti atoms with different coordination for the variously sized nanospheres, as optimized at DFTB level, in terms of their number and percentage with respect to the total number of Ti atoms, is reported. Corresponding values for the optimized nanosphere at the DFT(B3LYP) level are in parenthesis, only when different from DFTB.

| DFTB [DFT(B3LYP)] | Number | % | Number | % | Number | % | Number | % |
|---|---|---|---|---|---|---|---|---|
| Ti Site | 1.5 nm | | 2.2 nm | | 3.0 nm | | 4.4 nm | |
| $Ti_{4c}$ | 20 (19) | 19.8 (18.8) | 36 | 16.1 | 53 | 13.3 | 106 | 8.4 |
| $Ti_{5c}$ | 20 (21) | 19.8 (20.8) | 43 (49) | 19.2 (22.0) | 69 (65) | 17.3 (16.3) | 159 | 12.6 |
| $Ti_{6c\_sup}$ | 20 | 19.8 | 28 (24) | 12.6 (10.8) | 72 (75) | 18 (18.8) | 157 | 12.4 |
| $Ti_{6c}$ | 29 | 28.7 | 96 (94] | 43.1 (42.1) | 181 (182) | 43.4 (45.6) | 791 | 62.5 |
| Ti$_{3c}$(OH) | 8 | 7.9 | 8 | 3.6 | 16 | 4 | 20 | 1.6 |
| Ti$_{4c}$(OH) | 4 | 4 | 12 | 5.4 | 8 | 2 | 32 | 2.5 |

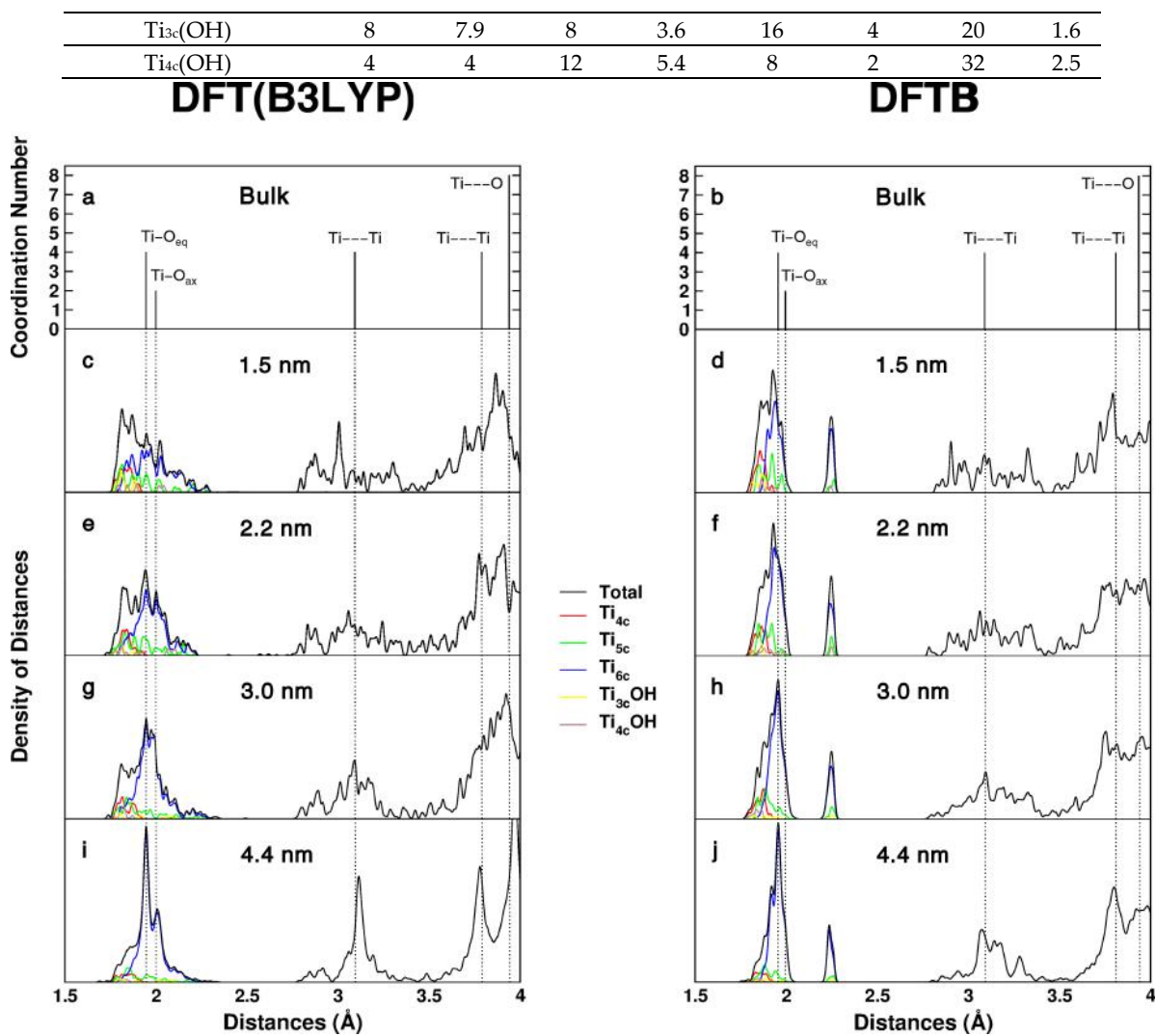

**Figure 2.** Distances distribution (simulated EXAFS) computed with DFT(B3LYP) and DFTB for bulk anatase (**a**,**b**), computed with DFT(B3LYP) for the: 1.5 nm (**c**); 2.2 nm (**e**); 3.0 nm (**g**); and 4.4 nm (**i**) nanospheres; and computed with DFTB for the: 1.5 nm (**d**); 2.2 nm (**f**); 3.0 nm (**h**); and 4.4 nm (**j**) nanospheres.

Hybrid functional B3LYP simulated EXAFS spectrum for the 2.2 nm nanoparticle was compared to the corresponding one obtained from the fully optimized 2.2 nm nanoparticle by the semilocal PBE functional in Figure S1. It is interesting to note that the two curves almost overlap, showing an excellent agreement between the two methods.

### 3.2. Electronic Properties

The electronic properties of a semiconducting oxide with a relatively large band gap as TiO$_2$ are not simply described by any quantum mechanical method. Standard DFT methods severely underestimate the band gap value, whereas hybrid DFT, as a consequence of the contribution of exact exchange in the exchange functional, provide values in closer agreement with experiments [22]. DFTB has been tested for bulk TiO$_2$ calculations and found to be in excellent agreement with experimental data and DFT Hubbard corrected values [23]. Therefore, DFTB is expected to perform well when investigating TiO$_2$ nanoparticles.

In the following, we present a comparison of the density of states (DOS) for the nanospheres of different size, as shown in Figure 3, that have been obtained with DFT(B3LYP) and DFTB on the corresponding fully relaxed structures. Considering that nanoparticles are finite systems, one could not really define true band states and band gaps. We have decided to distinguish between very localized states (molecular orbitals) and delocalized on several atoms of the nanoparticle (pseudo band states). These two definitions, based on a threshold value of 0.02 for the maximum squared coefficient of each eigenstate ($max_c$), lead to two different values of gap: the HOMO-LUMO gap and the BAND gap, which are reported in Table 2. We observe a decreasing trend with both DFT(B3LYP) and DFTB methods. The BAND gap values progressively approach the bulk Kohn–Sham value of 3.81 eV for DFT(B3LYP) and of 3.22 eV for DFTB (the experimental band gap for bulk anatase is 3.4 eV at 4 K) [52]. This is in line with experimental data [53] based on UV-Vis optical techniques the band gap of $TiO_2$ nanoparticles increases with decreasing size, due to quantum confinement effects.

**Table 2.** HOMO-LUMO electronic gap ($\Delta E_{H-L}$) and electronic BAND gap $\Delta E_{BAND}$ (expressed in eV) calculated for nanospheres of different size and for bulk anatase with both DFT(B3LYP) and DFTB methods.

| MODEL | $\Delta E_{H-L}$ | | $\Delta E_{BAND}$ | |
|:---:|:---:|:---:|:---:|:---:|
| **Nanospheres** | **DFT(B3LYP)** | **DFTB** | **DFT(B3LYP)** | **DFTB** |
| 1.5 nm | 4.23 | 3.12 | 4.81 | 3.62 |
| 2.2 nm | 4.13 | 3.11 | 4.31 | 3.55 |
| 3.0 nm | 4.00 | 2.95 | 4.13 | 3.42 |
| 4.4 nm | 3.92 | 2.95 | 3.96 | 3.33 |
| BULK | - | - | 3.81 | 3.22 |

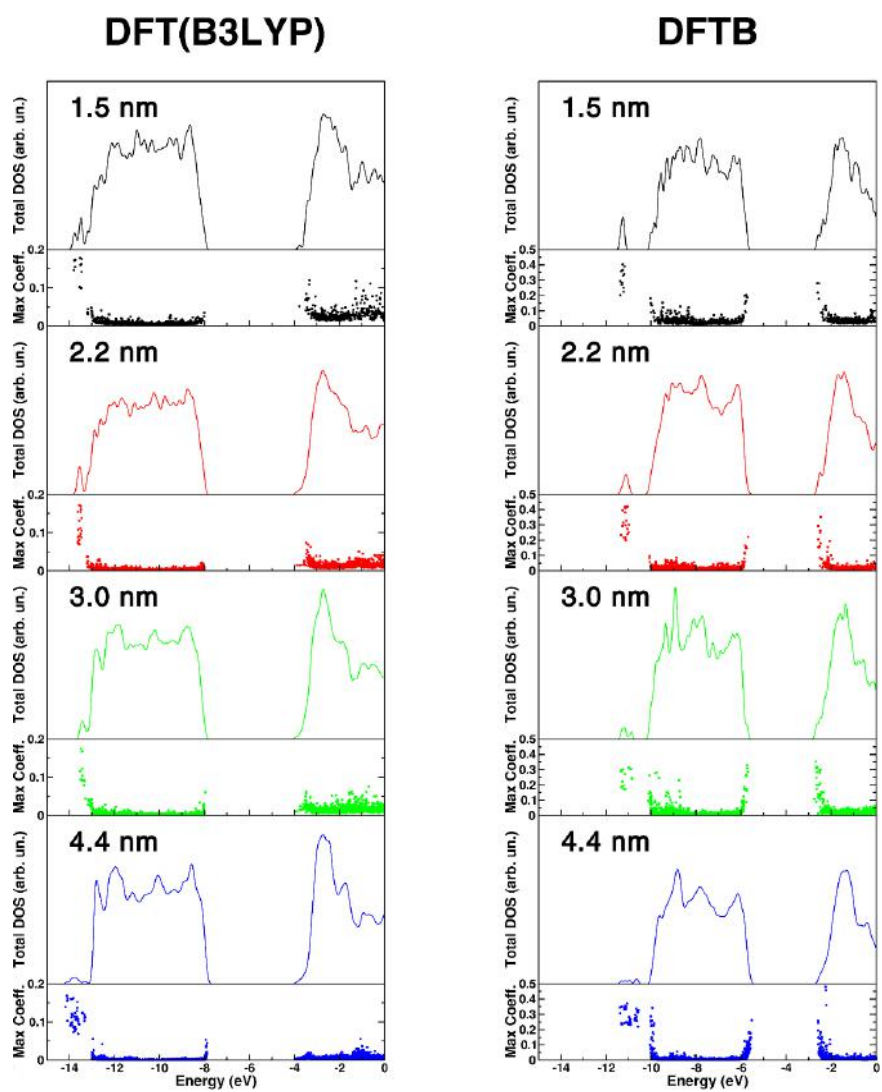

**Figure 3.** DFT(B3LYP) and DFTB total (DOS) density of states for different size nanosphere, 1.5 nm (black), 2.2 nm (red), 3.0 nm (green), and 4.4 nm (blue). For each nanosphere, the DOS has been normalized to the number of TiO$_2$ units to have comparable DOS intensities. The maximum atomic orbital coefficient ($max_c$) of each eigenstate is also reported. High values of $max_c$ correspond to localized states, while low values correspond to delocalized states.

The DOS curves shown in Figure 3 for the different nanospheres further confirm a band gap opening going from the largest one (bottom panel) to the smallest one (upper panel). Additionally, we present the value of $max_c$ for each eigenstate because we wish to highlight the degree of localization/delocalization of the states making up the DOS. The higher the value of $max_c$, the higher the localization. DFT(B3LYP) results show some localization at the band edges (top of the valence band and bottom of the conduction band) and in the range of OH groups in the low energy range (at about −14 eV). DFTB results are qualitatively similar, although some excess localization can be observed. In Figure S2 of the Supplementary Materials we have also reported and compared the total DOS for the anatase bulk TiO$_2$ calculated at DFT(B3LYP) and DFTB level of theory.

We may conclude this section devoted to the preparation and description of models for TiO$_2$ spherical nanoparticles with the following summarizing remarks: spherical models carved from bulk supercells must be made chemically stable by the introduction of some hydroxyl groups that saturate highly undercoordinated sites; such rough models must then undergo a simulated thermal annealing (with DFTB method) that allows to achieve global minimum stable structures; those must be then further optimized either with DFTB or, to reach even higher accuracy, with a hybrid DFT method (here B3LYP). Although some fine details are different, the general picture we obtain with DFTB is rather similar to that from DFT(B3LYP), which assesses DFTB as a reliable method to investigate nanoparticles of large realistic size (up to 4000 atoms, corresponding to a diameter of 4.4 nm). This will allow for further future development and for the study of nanoparticles' surface functionalization.

## 4. Modelling Photoactivation of TiO$_2$ Nanoparticles

Titanium dioxide is still considered the reference system in the research fields of photocatalysis and photovoltaics for its ability to convert light photons into chemical energy. In the very beginning of the photocatalytic process in a semiconductor, when the material is irradiated with light, an electron/hole pair or "exciton", is initially formed [54,55]. Then, if the coupling with lattice vibrations is strong enough, the exciton may become self-trapped (self-trapped exciton, STE) on few atoms of the crystal, reducing significantly its mobility. Finally, the photoexcited charge carriers may: (i) migrate towards the surface of the semiconductor as trapped electrons or holes and express their intrinsic redox activity; or (ii) recombine radiatively and emit a photoluminescence photon.

In this regard, the quantum confinement of an exciton in a nanoparticle with a dimension of few nanometers may significantly influence its size and localization [56]. Furthermore, in a TiO$_2$ spherical nanoparticle, the close presence of highly undercoordinated surface sites may be a driving force for the process of excitons separation into electrons and holes or, on the contrary, may accelerate the radiative recombination via exciton self-trapping processes.

The study of photoexcited charge carriers in nanoparticles, although very demanding, must be performed by using a hybrid functional method (here B3LYP), since any other local or semilocal functional, and therefore also DFTB, would not properly describe the degree of electron/hole localization as a consequence of the self-interaction problem inherent in those methods [57]. DFT+U approach could be an alternative viable route; however, hole trapping was found to required very high and unphysical U values [58].

### 4.1. Free/Trapped Excitons and Radiative Recombination

We first recall the nature of the self-trapped exciton (STE) in bulk anatase, as shown in Figure 4 and reported in more detail in a previous work [27]. When the system is allowed to relax from the fully delocalized exciton initial condition (Figure 4b), two different self-trapped excitons can be localized, which differ for the O atom of the TiO$_6$ octahedron involved in the trapping: this can be either the equatorial O with respect to the electron trapping Ti (Ti$^{3+}$-O$_{eq}^-$ in Figure 4c) or the axial one (Ti$^{3+}$-O$_{ax}^-$ in Figure 4d).

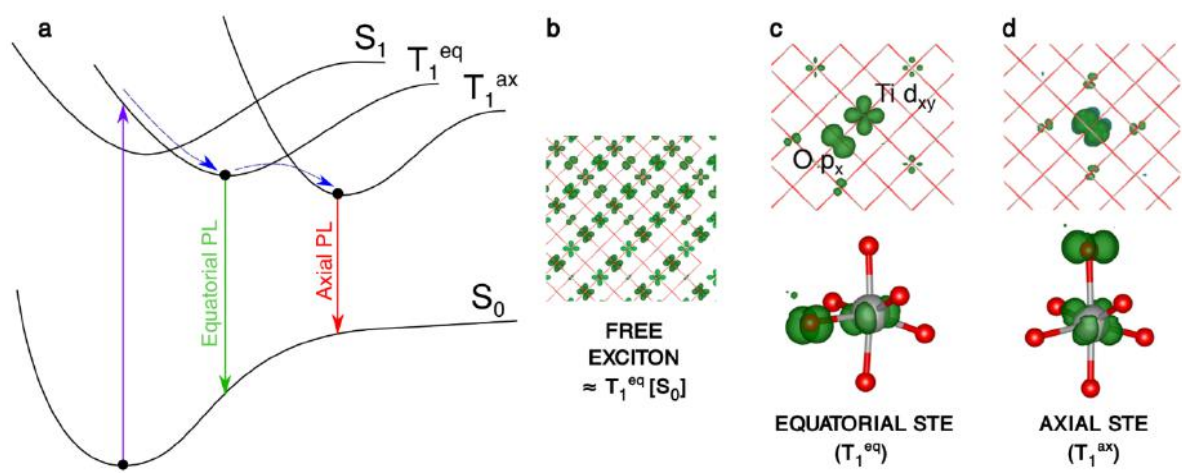

**Figure 4.** (**a**) Schematic representation of the processes involving the electron/hole couple: the vertical excitation S$_0$ → T$_1$, the self-trapping relaxation in the bulk structure and the two different photoemissions T$_1^{eq}$ → S$_0$ and T$_1^{ax}$ → S$_0$. (**b–d**) 3D spin density plots of the anatase bulk supercell, as obtained with the B3LYP functional, for the vertical triplet state (**b**), trapped triplet equatorial (**c**) and axial (**d**) exciton in the bulk. The spin density isovalue is 0.01 a.u. (0.0005 a.u. for the vertical triplet).

The trapping energy (ΔE$_{trap}$), that is defined here and in the following as the energy difference between the fully relaxed trapped system state and the fully delocalized solution in the system ground state geometry, is more negative by −0.1 eV for the axial exciton (see Table 3). The computed photoluminescence (PL) energies for the decay of these two self-trapped excitons are given in Table 3. Noteworthy, the average PL is 2.24 eV, in very good agreement with the experimental value of 2.3 eV [59].

**Table 3.** Trapping Energy (ΔE$_{trap}$) of the exciton in the triplet state and the corresponding photoluminescence (PL) in the axial and equatorial configuration for bulk anatase and for the nanosphere. All energies are in eV.

|  |  | **Bulk** | **Nanosphere** |
|---|---|---|---|
| Ti$^{3+}$-O$_{ax}^-$ | ΔE$_{trap}$ | −0.59 | −0.65 |
|  | PL | 1.99 | 1.96 |
| Ti$^{3+}$-O$_{eq}^-$ | ΔE$_{trap}$ | −0.49 |  |
|  | PL | 2.35 |  |

In the spherical anatase nanoparticle model, the vertical excitation does not lead to a fully delocalized solution as for bulk (Figure 5a), but the resulting exciton partially localizes on a portion of the curved surface, even if the nanosphere is in its ground state optimized structure. After atomic relaxation, the exciton becomes trapped in the core of the nanosphere (Figure 5b). As for bulk, the axial solution is favored, with trapping and photoluminescence energies similar to the bulk values (Table 3). Thus, to summarize, a confinement effect in the nanosized systems is observed for the "Franck–Condon" exciton, but not for the self-trapped exciton in the core.

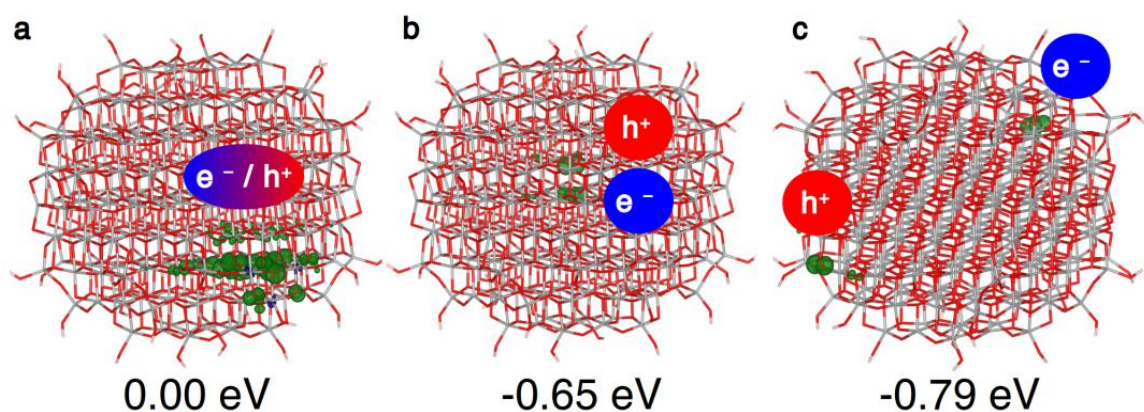

**Figure 5.** 3D spin density plots of the spherical $TiO_2$ nanoparticle, as obtained with the B3LYP functional, for the vertical triplet state (**a**), trapped triplet exciton (**b**) and state (**c**) with the hole and the electron at the best trapping sites. The spin density isovalue is 0.01 a.u. (0.002 a.u. for the vertical triplet).

As a next step, if the electron/hole couple has enough energy to separate, the charge carriers may migrate to the surface, where many trapping sites are available. Indeed, as we will discuss in Section 4.2, the most stable configuration for the electron and the hole are a subsurface $Ti_{6c}$ site and a surface $O_{2c}$ site, respectively. Considering these as trapping sites for the electron and the hole, the trapping energy of the separated exciton, shown in Figure 5c, amounts to −0.79 eV, significantly larger than the one of the bound exciton in the core (−0.65 eV). Therefore, there is a favorable energy gradient for the electrons and holes to separate and to move towards the surface, which is the driving force for separation and migration processes.

### 4.2. Separated Carriers Trapping

When the charge carriers are far apart in different regions of the nanosphere, as in Figure 5c, they behave almost like isolated charges. Thus, it is possible to study the relative stability of electron or hole trapping sites introducing a single extra electron or extra hole in the system [28].

Within an anatase spherical nanoparticle, we observed that the excess electron, when added to the system in its ground state geometry, fully delocalizes on all the Ti centers (see Figure 6a), except those in the outermost atoms on the curved surface. This is used as the reference system for the evaluation of the trapping energy ($\Delta E_{trap}$).

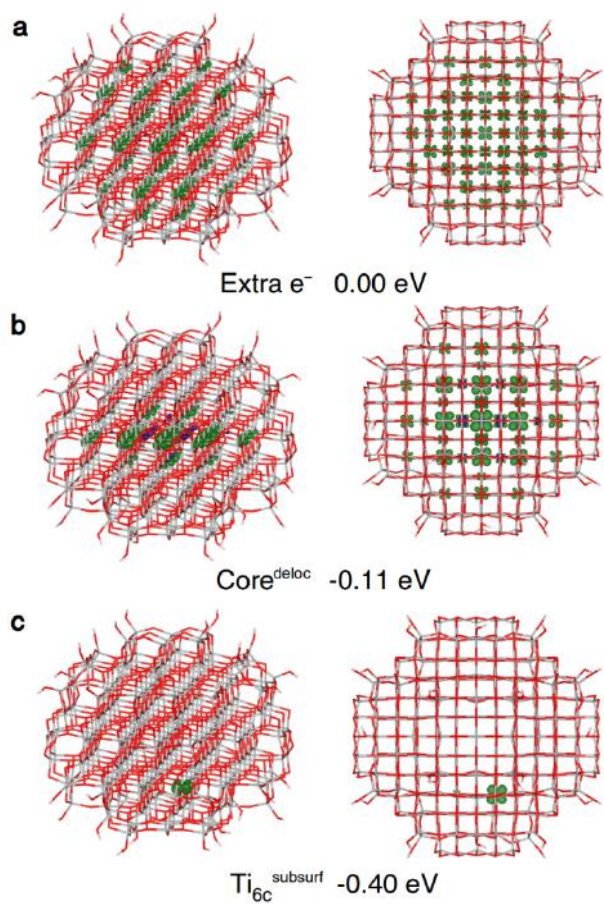

**Figure 6.** Front and top view of the 3D plots of spin density of: (**a**) an extra electron added to the neutral ground state structure (isovalue = 0.001 a.u.); (**b**) a trapped delocalized electron (isovalue = 0.005 a.u.); and (**c**) an electron trapped on a subsurface six-fold coordinated titanium atom (isovalue = 0.001 a.u.) in the anatase nanosphere model, as obtained with the B3LYP functional. Below each structure, the energy gain ($\Delta E_{trap}$) relative to (**a**) is given.

After atomic relaxation in the presence of this extra electron, the spin density localizes on several Ti atoms within the central three atomic core layers (see Figure 6b), with an energy gain of −0.11 eV. In other words, a quite stable large polaron, involving few atomic layers, is formed. This partially trapped intermediate situation, also referred to as "shallow trap" [3], cannot be observed in periodic models of bulk and slabs.

We also investigated the electron localization at "deep traps", i.e., single atomic Ti sites in the nanosphere. The trapping energy and the electron localization are shown in Table 4. Unexpectedly, no trapping has been observed on undercoordinated Ti atoms, i.e., four-fold coordinated at the equator of the nanosphere (Ti$_{4c}$equator), four-fold coordinated Ti with a terminal OH (Ti$_{4c}$(OH)) and five-fold coordinated (Ti$_{5c}$) sites, except for a very small trapping energy (−0.09 eV) in the case of the Ti$_{4c}$equator site (see Table 4).

On the contrary, the best electron trap is the fully coordinated Ti$_{6c}$ site on the subsurface (Ti$_{6c}$subsurf in Table 4 and Figure 6c) with a $\Delta E_{trap}$ of −0.40 eV. Noteworthy, the electron delocalization in the core of the nanoparticle (Core$^{deloc}$ in Table 4 and Figure 6b) has been found to be less favored than complete or full localization on a single subsurface Ti site (−0.11 vs. −0.40 eV), indicating an energy gradient, and thus a driving force, for the migration and localization of photoexcited electrons towards atoms near the surface.

**Table 4.** Trapping energy ($\Delta E_{trap}$) for electrons at different sites in the spherical anatase nanoparticles, as obtained with B3LYP functional. The reference zero for $\Delta E_{trap}$ is obtained by adding one electron to the nanosphere in its neutral ground state geometry, with no atomic relaxation. The charge localization (%electron) is also given. The sites nomenclature is defined in the text.

| Position | $\Delta E_{trap}$ (eV) | %electron |
|---|---|---|
| $Ti_{4c}^{equator}$ | −0.09 | 88% |
| $Ti_{4c}(OH)$ | No trapping | |
| $Ti_{5c}$ | No trapping | |
| $Ti_{6c}^{subsurf}$ | −0.40 | 85% |
| $Ti_{6c}^{core}$ | −0.13 | 62% |
| $Core^{deloc}$ | −0.11 | 19% |

Concerning the hole trapping on spherical TiO$_2$ nanoparticles, if one electron is removed without any atomic relaxation (vertical ionization), only some regions of the nanoparticles are involved, as shown in Figure 7a. Since this solution cannot be seen as a delocalized band-like state of a free (or untrapped) hole, its total energy cannot be used as the reference to compute trapping energies ($\Delta E_{trap}$). Thus, in the following we will use adiabatic ionization potentials (IPs) for comparisons, since they correlate with trapping energies: the smaller the adiabatic IP, the larger the trapping energy of the considered site. In the core of the nanosphere, the hole almost completely localizes (90%) on the central three-fold coordinated O atom ($O_{3c}^{core\_ax}$ in Figure 7b and Table 5). Differently from what found for electrons and discussed above, we could not identify any delocalized "shallow trap" state for holes.

If we allow the hole to reach the curved surface of the nanosphere, several types of O sites are available for trapping. Among them, the most stable one is a two-fold O atom bridging a $Ti_{5c}$ and a $Ti_{6c}$ atom ($O_{2c}^{5c-6c}$ Table 5 and Figure 7c). A second type of two-fold O on the surface, between a $Ti_{4c}$ and a $Ti_{6c}$ atom (see $O_{2c}^{4c-6c}$ in Table 5), can also trap the hole but less efficiently than a $O_{2c}^{5c-6c}$ site. It is worth underlining that the migration of holes from the nanosphere core to the surface is energetically favourable by −0.12 eV.

Finally, one should note that on the surface of a spherical nanoparticle there are several hydroxyl groups that may be stable hole trapping sites. However, as reported in Table 5, we were not able to localize the hole on the hydroxyl group of a $Ti_{3c}(OH)$ and $Ti_{4c}(OH)$ sites on the nanosphere surface. On the contrary, we could trap the hole on a $Ti_{5c}(OH)$ site, formed upon dissociation of a water molecule on a $Ti_{5c}$, probably because a OH bound to a fully coordinated Ti site is more electron rich than one bound to an undercoordinated Ti atom. Nonetheless, the OH trapping site is less effective by 0.29 eV than the most stable $O_{2c}$ hole trapping site. Therefore, in vacuum, the OH groups are not good trapping sites, but the scenario may change in an aqueous medium, where water molecules may enhance the trapping properties of the hydroxyl groups, through binding as a ligand to TiOH or H-bonding to the hydrogen of the OH.

**Table 5.** Adiabatic ionization potential (IP$^{ad}$) for holes at different sites of the spherical anatase nanoparticle, as obtained with B3LYP functional. The charge localization (%hole) is also reported. The sites nomenclature is defined in the text.

| Position | IP$^{ad}$ (eV) | %hole |
|---|---|---|
| $O_{2c}^{4c-6c}$ | 7.74 | 89%/11% |
| $O_{2c}^{5c-6c}$ | 7.52 | 92%/8% |
| $O_{3c}^{core\_ax}$ | 7.68 | 90% |
| $Ti_{3c}(OH)$ | No trapping | |
| $Ti_{4c}(OH)$ | No trapping | |
| $Ti_{5c}(OH)$ | 7.81 | 85%/15% |

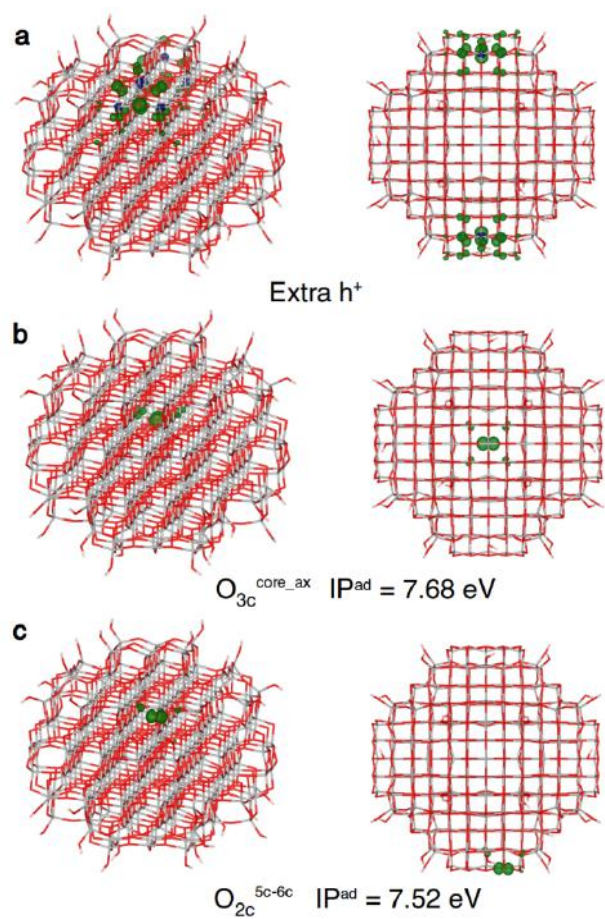

**Figure 7.** Front and top view of the 3D plots of spin density of: (**a**) an excess hole resulting from a vertical ionization of the model (isovalue = 0.001 a.u.); (**b**) a trapped hole in the core (isovalue = 0.01 a.u.); and (**c**) a hole trapped on a two-fold coordinated oxygen atom (isovalue = 0.01 a.u.) in the anatase nanosphere model, as obtained with the B3LYP functional. Below the structures with a trapped hole the adiabatic ionization potential (IP$^{ad}$) as a measure of the hole trapping ability is given.

### 4.3. Comparison with Experiments

Experimental data on trapped charges in anatase TiO$_2$ are available in literature and they can be compared with calculations performed with the spherical nanoparticle models shown above. First, the calculated values of the trapping energy relative to a free electron in the conduction band are in good agreement with the experimental observations for both shallow (delocalized) [60–63] and deep (localized) [64–66] electron trapping states.

Moreover, the degree of electron localization can be probed through the hyperfine coupling constant (a$_{iso}$) with the next-neighboring $^{17}$O in the electron paramagnetic resonance (EPR) spectrum. High values of a$_{iso}$ are expected for localized electrons, low values for delocalized ones. Indeed, the computed a$_{iso}$ is 6.7 MHz for an electron localized on the innermost Ti atom of the NP (see Ti$_{6c}^{core}$ Table 4), whereas it is 3.9 MHz for an electron delocalized in the NP core (Core$^{deloc}$ in Table 4), in good agreement with experimental observations of a significant decrease of a$_{iso}$ going from a fully localized electron on a single Ti ion (as in the Ti$^{3+}$(H$_2$O)$_6$ complex) [67] to shallow electron traps in anatase nanoparticles [68]. For the most stable hole trap on the surface, the computed EPR parameters (g = [2.004, 2.015, 2.019] G and A = [31, 30, −97] G) are in excellent quantitative agreement with the g- and A-tensor data available in the experimental literature [69,70]. Hence, we may conclude that a correct localization of both charge carriers is provided by the computational models and methods.

Finally, we employed the transition level approach [27,71] to estimate the electronic transition energies of the charged traps, since this methodology produces accurate results for excitations of defects in solids. In the case of electrons, the calculated transition from the trap level of the best electron trap ($Ti_{6c}^{subsurf}$ in Table 4) to the conduction band minimum (CBM) is 1.25 eV, in accordance with the experimental value of 1.37 eV, measured with transient absorption (TA) spectroscopy [72]. In the case of holes, we computed a transition from the valence band maximum to the trap level for the best hole trapping site ($O_{2c}^{5c-6c}$ in Table 5) of 2.59 eV, to be compared with a reported experimental value of 1.9 eV in the experimental TA spectrum [72]. This inconsistency between the computed and experimental results may arise because these experiments have been performed in an aqueous solution and, as mentioned in Section 4.2, the presence of water layers on the nanoparticle may influence the hole trapping ability of the system, as reported in a recent experimental work [35].

To conclude this section devoted to the study of the life path of energy carriers (excitons) and charge carriers (electron and holes) in spherical $TiO_2$ nanoparticles by hybrid DFT(B3LYP), we can summarize as follows: the photoexcited exciton self-trapping is a favorable process but electron and hole can then separate to migrate towards the surface where they can be highly stabilized. In particular, for electrons, we observed that deep trapping at subsurface fully coordinated Ti sites is favored with respect to shallow trapping in the core. In the case of holes, only deep traps were observed with a surface $O_{2c}$ ($Ti_{5c}$-O-$Ti_{6c}$) being the preferential hole trapping site. Computed electron paramagnetic resonance (EPR) parameters and optical transitions for those electron/hole traps are in good agreement with experimental data.

## 5. Modelling Surface Interaction with Water

Understanding the interaction of water with $TiO_2$ anatase surface [73] is essential since $TiO_2$-based technologies, including photocatalysis, normally operate in an aqueous environment. Many computational studies based on DFT methods have tackled the interaction between the most exposed anatase $TiO_2$ (101) surface and water layers [74–76] revealing how the surface complexity influences the water-titania interface.

However, the study of the dynamical behavior of real size $TiO_2$ nanoparticles (i.e., with diameter in the range 2–8 nm) [53,69,77–80] in a realistic aqueous environment and with sufficiently long simulation times, is currently not feasible with DFT methods.

As regards DFTB, from a technical point of view, its performance in the description of a certain system critically depends on the parameterization of the element-pairs interaction of the atoms involved. In the case of Ti-containing compounds, two different sets of parameters are available: "mio-1-1/tiorg-0-1" [23] and "matsci-0-3" [24]. The first set has been developed to handle the interaction of low index surfaces of both anatase and rutile with water and small organic molecule, but no assessment has been done for the anatase $TiO_2$ (101) surface. The second set has been thought to describe bulk $TiO_2$ structures and chemical reactivity of (101) anatase and (110) rutile surfaces with isolated molecule and monolayers of water. However, this set has been never tested for the description of bulk water, which is essential for a correct characterization of titania/water-multilayers interfaces.

Recently, we have shown [40] that if these two sets are properly combined in a new one, referred to as "MATORG", with some further improvement coming from the inclusion of an empirical correction [51] for a finer description of the hydrogen bonding ("MATORG+HBD"), it is possible to achieve a DFT-like description of the interaction between water-multilayers and anatase $TiO_2$ (101) surface. In the following, we will shortly present the performance of this new set of parameters for the static and dynamic description of $TiO_2$/water interface by comparison with previous DFT results. The positive assessment of DFTB methods for studying this type of solid/liquid interface is extremely important because it gives a solid basis for its application on large realistic nanoparticles in an aqueous environment.

*5.1. Bulk Water and Anatase TiO2 Description*

The correct description of the titania/water-multilayers interface is tightly related to the method ability of properly describing the two components separately. Bulk TiO₂ lattice parameters have been calculated with DFTB and compared, in Table 6, with DFT and experimental values reported in literature [17,47,81,82]. The agreement is extremely good. The MATORG+HBD set gives an extremely accurate *a* value and only slightly overestimates the *c* lattice parameter and consequently the *c*/*a* ratio.

**Table 6.** Lattice parameters *a* and *c* and their *c*/*a* ratio for bulk TiO₂ anatase. Values computed with the DFTB and DFT methods are reported and compared with the experimental values. In parenthesis, the absolute errors referred to the experimental data are shown.

| Method | Reference | *a* (Å) | *c* (Å) | *c*/*a* (Å) |
|---|---|---|---|---|
| **DFTB-MATORG+HBD** | This work | 3.796 (+0.014) | 9.790 (+0.288) | 2.579 (+0.067) |
| **DFT(PBE)** | This work | 3.789 (+0.007) | 9.612 (+0.110) | 2.537 (+0.025) |
| **DFT(PBE)** | Ref. [81] | 3.786 (+0.004) | 9.737 (+0.235) | 2.572 (+0.060) |
| **DFT(B3LYP)** | Ref. [82] | 3.783 (+0.001) | 9.805 (+0.303) | 2.592 (+0.080) |
| **DFT(B3LYP)** | Ref. [17] | 3.789 (+0.007) | 9.777 (+0.275) | 2.580 (+0.068) |
| **DFT(HSE06)** | Ref. [17] | 3.766 (-0.016) | 9.663 (+0.161) | 2.566 (+0.054) |
| **Exp.** | Ref. [47] | 3.782 | 9.502 | 2.512 |

To assess the reliability of DFTB method and MATORG+HBD parameters for the description of bulk water, two different features are evaluated: the H-bond strength and the radial distribution function (RDF). We estimate the H-bond strength ($\Delta E_{H-bond}$) by the water dimer binding energy and report it in Table 7, together with the equilibrium O-O distances ($R_{O-O}$) of the water dimer. DFTB results are compared with those obtained with standard and hybrid DFT [83], CCSD [84] and experiments [85,86]. The description of the H-bond with the MATORG+HBD set is extremely good, very close to the first-principles and experimental references.

**Table 7.** Water dimer binding energy ($\Delta E_{H-bond}$) and oxygen–oxygen distance ($R_{O-O}$). Values obtained with DFTB (MATORG+HBD), higher-level methods (CCSD, PBE and B3LYP) and experiments are reported. In parenthesis, the absolute errors referred to the experimental data are shown.

| Method | Reference | $\Delta E_{H-bond}$ (eV) | $R_{O-O}$ (Å) |
|---|---|---|---|
| **DFTB-MATORG+HBD** | This work | 0.199 (−0.037) | 2.815 (−0.157) |
| **DFT(PBE)** | Ref. [83] | 0.222 | 2.889 |
| **DFT(B3LYP)** | Ref. [83] | 0.198 | 2.926 |
| **CCSD** | Ref. [84] | 0.218 | 2.912 |
| **Exp.** | Refs. [85,86] | 0.236 | 2.972 |

The radial distribution functions (RDF) of oxygen–oxygen ($O_w$-$O_w$) and hydrogen–hydrogen ($H_w$-$H_w$) by DFTB with the MATORG+HBD set of parameters are compared with the experimental ones in Figure 8 [87].

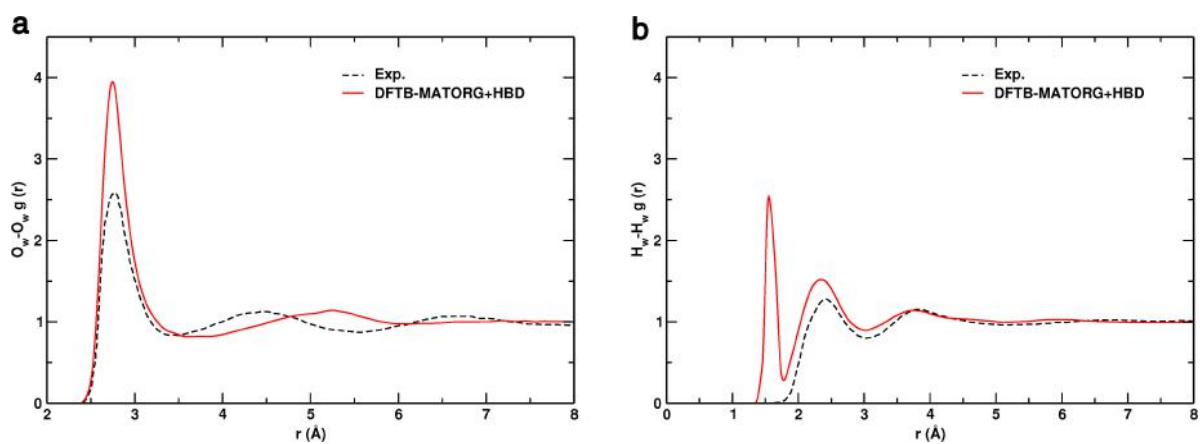

**Figure 8.** Comparison of the: oxygen–oxygen ($O_w$-$O_w$) (**a**); and hydrogen–hydrogen ($H_w$-$H_w$) (**b**) radial distribution functions (RDF) obtained experimentally (dashed black line) and calculated with the DFTB-MATORG+HBD method.

In the experiment, the first intermolecular peaks for r($O_w$-$O_w$) and r($H_w$-$H_w$) are found to be located at 2.77 and 2.31 Å, respectively. In excellent agreement, with the MATORG+HBD set, the first peak position of the $O_w$-$O_w$ RDF is at 2.75 Å (Figure 8a). The experimental/theoretical curves partially overlap for distances lower than 3.30 Å: the water density depletion zones are very similar and the second intermolecular peak is at 5.25 Å, not too far from the experimental data (4.55 Å). Regarding the $H_w$-$H_w$ radial distribution function (Figure 8b), the experimental curve is well reproduced by MATORG+HBD, with the first intermolecular peak located at 2.34 Å.

### 5.2. Static Description of TiO₂/Water Interface

The molecular (undissociated, $H_2O$) and dissociated (OH, H) adsorption of water on the anatase TiO₂ (101) surface has been investigated at different water coverages (low, $\theta = 0.25$ and full, $\theta = 1$). The adsorption energy per molecule ($\Delta E_{ads}^{mol}$) has been calculated with MATORG and MATORG+HBD and compared with DFT(PBE) results and experimental measurements in Table 8.

**Table 8.** Values calculated with DFT and DFTB methods of the adsorption energies per molecule ($\Delta E_{ads}^{mol}$) of water on the TiO₂ (101) anatase slab in the molecular ($H_2O$) and dissociated (OH, H) state. Different coverages are considered (low, $\theta = 0.25$ and full, $\theta = 1$). The experimental adsorption energy of the water monolayer on the (101) surface is also reported. The absolute errors (in parenthesis) reported for DFTB are calculated with respect to the PBE values from this work.

| Method | Reference | Coverage, $\theta$ | $\Delta E_{ads}^{mol}$, $H_2O$ (eV) | $\Delta E_{ads}^{mol}$, OH,H (eV) |
|---|---|---|---|---|
| **DFTB-MATORG** | This work | 0.25 | −1.08 (+0.41) | −0.54 (+0.22) |
| | | 1 | −0.96 (+0.34) | −0.58 (+0.15) |
| **DFTB-MATORG+HBD** | This work | 0.25 | −0.80 (+0.13) | −0.31 (−0.01) |
| | | 1 | −0.71 (+0.09) | −0.40 (−0.03) |
| **DFT(PBE)** | This work | 0.25 | −0.67 | −0.32 |
| | | 1 | −0.62 | −0.43 |
| **DFT(PBE)** | Ref. [88] | 0.25 | −0.74 | −0.23 |
| | | 1 | −0.72 | −0.44 |
| **Exp.** | Refs. [89.90] | 1 | −0.5/−0.7 | |

The DFTB method predicts the molecular adsorption mode of a single water molecule to be favored with respect to the dissociated one. This is in line with several experimental observations [73,89–91] and previous DFT data [10,88]. In the full coverage regime ($\theta = 1$), the MATORG set also correctly reproduces the binding energy decrease for the molecular adsorption mode and the increase for the dissociated one. However, the MATORG set tends to overestimate adsorption energies with errors up to 0.41 eV. This discrepancy is almost solved with the inclusion of the HBD correction, which reduces the error values to less than 0.13 eV.

### 5.3. Dynamic Description of TiO₂/Water Interface

The study of complex and realistic TiO₂ (nano)systems in aqueous environment is strongly related to the ability of the method used to describe the titania/water-multilayers dynamic behavior. To assess the performance of the MATORG+HBD set of parameters, first-principles simulations must be used as reference. We will use Car–Parrinello molecular dynamics (CPMD) DFT(PBE) simulations [75] and other DFT(PBE) structural investigations [92] of water layers on the TiO₂ (101) anatase surface, which already exist in the literature.

In Figure 9, we show the 0 K optimized geometries starting from the last snapshot of each molecular dynamics trajectory performed with the MATORG+HBD set, in the case of the fully undissociated water monolayer (ML), bilayer (BL) and trilayer (TL) of water on the (101) TiO₂ anatase surface, respectively.

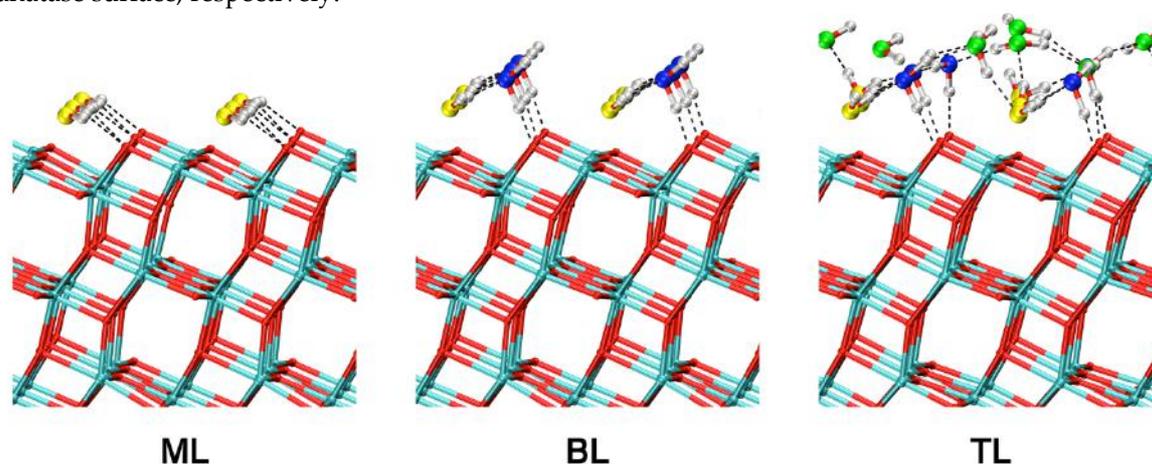

**Figure 9.** DFTB-MATORG+HBD structures of a monolayer (ML), a bilayer (BL) and a trilayer (TL) of water on the (101) TiO₂ anatase surface. Dashed lines correspond to H-bonds.

For the ML, each water molecule (in yellow in Figure 9, left panel) binds a Ti₅c of the surface and establishes two H-bonds with the O₂c with an adsorption energy per molecule ($\Delta E_{ads}^{mol}$ in Table 9), calculated with the MATORG+HBD method, of −0.70 eV, in very good quantitative agreement with the DFT(PBE) references [74].

In the BL configuration, the water molecules of the first layer are bound to Ti₅c atoms of the surface and form two H-bonds with two molecules of the second layer. The water molecules of the second layer (in blue in Figure 9, middle panel) have only one H-bond with an O₂c of the titania surface, with the other H atom pointing towards the vacuum. For the BL, the MATORG+HBD adsorption energy is −0.73 eV, in agreement with DFT(PBE) results (see Table 9).

The TL case is more complicated, since the third water layer (in green in Figure 9, right panel) is too mobile to allow for a unique structure definition. We added a third water layer on the BL equilibrium structure with a MATORG+HBD adsorption energy of −0.53 eV (see Table 9), again in very good agreement with the DFT(PBE) previous study.

**Table 9.** Values calculated with DFT and DFTB methods of the binding energy per molecule ($\Delta E_{ads}^{mol}$ in eV) of the water monolayer (ML), bilayer (BL) and trilayer (TL) on the (101) TiO₂ anatase surface after an optimization run from the last snapshot of the MD simulation. The binding energy ($\Delta E_{ads}^{mol}$) is defined as the difference between the total energy of the titania/water interface equilibrium structure and the sum of the total energy of six isolated water molecules plus the total energy of the optimized slab with one water layer less.

| Water Configuration | $\Delta E_{ads}^{mol}$ (eV) | | |
|---|---|---|---|
| | DFTB-MATORG+HBD | DFT(PBE) [a] | DFT(PBE) [b] |
| ML | −0.70 | −0.62 | −0.69 |
| BL | −0.73 | −0.67 | −0.65 |
| TL | −0.53 | −0.53 | −0.56 |

[a] This work; [b] from Ref. [74].

Finally, to analyze the behavior of the titania/water-multilayers interfaces during the MD simulations, the distribution $p(z)$ of the vertical distances between the O atoms of the $H_2O$ molecules and the $Ti_{5c}$ plane of the surface, together with their time evolution ($z(t)$), were extracted from the MD trajectory, as shown in Figure 10, and compared to DFT(PBE) CPMD results [75].

In the case of the ML (Figure 10, top panel), the agreement with the Car–Parrinello (PBE) molecular dynamics data is satisfactory: as it can be seen from the time evolution of perpendicular distances, the molecules librate around their equilibrium site and give a total $p(z)$ distribution very similar to the reference with the peak shifted by only 0.1 Å to shorter values.

Regarding the BL molecular dynamics simulation (Figure 10, central panel), the agreement with the CPMD (PBE) is extremely good. In the CPMD DFT(PBE) case, the position of the $p(z)$ distribution peak is at 2.15 Å for the first water layer and at 2.98 Å for the second one, whereas in the MD with MATORG+HBD, those are at 2.11 Å and at 3.08 Å, respectively. The BL configuration is very stable since none of the water molecule has left its initial equilibrium position in the whole simulation time.

For the water TL (Figure 10, bottom panel) we observe again a very good agreement between the DFT(PBE) and DFTB curves: the first two water layers in the MATORG+HBD molecular dynamics are vertically ordered in their initial equilibrium. On the contrary, the third layer water molecules are very mobile interacting with the second layer through H-bond. The range of the third layer vertical distances evaluated with MATORG+HBD is 3.6 < $z$ < 5.1 Å, thus shorter than the one calculated with DFT(PBE) (~ 4 < $z$ < 6 Å).

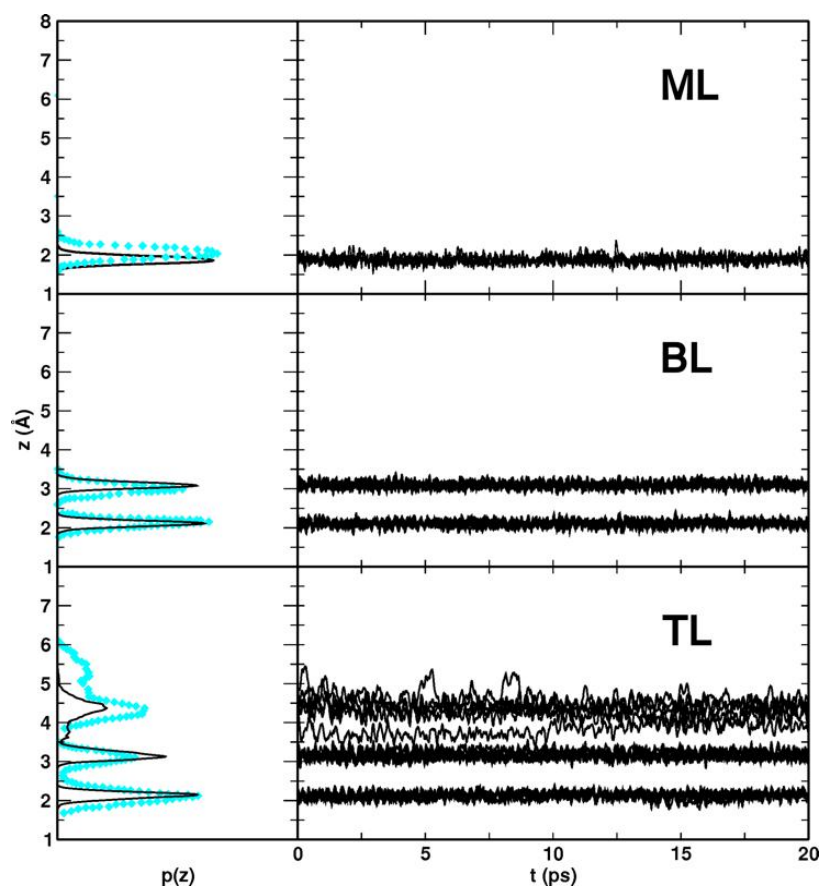

**Figure 10.** DFTB-MATORG+HBD distribution $p(z)$ and time evolution $z(t)$ of the distances between the water molecules (O atoms) of the monolayer (ML), bilayer (BL) and trilayer (TL) and the titania surface (Ti$_{5c}$ atoms). In cyan diamonds, values calculated with DFT(PBE) Car–Parrinello simulations are shown.

To conclude this section, we have shown that the description by the parametrized DFTB method with the MATORG+HBD set of the titania/water-multilayers interface (static and dynamic calculations) is in very good agreement with DFT(PBE) results. In particular, the MATORG+HBD set correctly describes the key aspects of the multilayer water adsorption on TiO$_2$ surface, properly balancing the surface/water and water/water interactions. Based on this assessment and on the computational efficiency of DFTB, we conclude that this method enables the study of large and realistic TiO$_2$ nanostructured systems into an aqueous environment.

## 6. Concluding Remarks

In the sections above, we have presented an overview of current possibilities, as explored by our group with state-of-the-art DFT and DFT-based (DFTB) methodologies, for the description of realistic nanoparticles in water solution for photoapplications. It is evident that, when the size of the nanoparticle becomes relatively large (about 4000 atoms), to achieve a diameter size close to the smallest TiO$_2$ nanoparticles used in practical applications (4.4 nm), the feasible limit for hybrid DFT calculations is almost reached. We showed that DFTB method can be used to obtain global minimum structures and to provide a reasonable description of both structural and electronic properties of these complex systems. Additionally, DFTB method is found to also yield a satisfactory accuracy for the description of the water layers on top of TiO$_2$ surfaces, which allows introducing the water environment explicitly into the calculations. However, we have shown that the DFT level of theory is still mandatory when one wants to describe photoexcitation processes taking place in the nanoparticles or on its surface, such as exciton formation, charge carrier trapping or electron transfer to adsorbates.

**Supplementary Materials:** The following are available online at www.mdpi.com/link, Figure S1: Comparison of the distances distribution (simulated EXAFS) computed with DFT(PBE), in black and DFT(B3LYP) in red, for the 2.2 nm NS produced at 300 K., Figure S2: DFT(B3LYP) and DFTB total (DOS) density of states for anatase bulk TiO$_2$. The maximum atomic orbital coefficient ($max_c$) of each eigenstate is also reported.

**Acknowledgments:** The authors are grateful to Prof. Gotthard Seifert for very fruitful discussions and to Lorenzo Ferraro for his technical help. The project has received funding from the European Research Council (ERC) under the European Union's HORIZON2020 research and innovation programme (ERC Grant Agreement No [647020]) and from CINECA supercomputing center through the computing LI05p_GRV4CUPT grant.

**Author Contributions:** G.F., D.S. and C.D.V. conceived the models and designed the calculations; D.S and G.F. performed the calculations; D.S., G.F. and C.D.V. performed the data analysis; C.D.V. wrote the manuscript with the help of D.S. and G.F.

**Conflicts of interest**: The authors declare no conflict of interest.

## References

1.　Sang, L.; Zhao, Y.; Burda, C. TiO$_2$ Nanoparticles as Functional Building Blocks. *Chem. Rev.* **2014**, *114*, 9283–9318.

2.　Ma, Y.; Wang, X.L.; Jia, Y.S.; Chen, X.B.; Han, H.X.; Li, C. Titanium Dioxide-Based Nanomaterials for Photocatalytic Fuel Generations. *Chem. Rev.* **2014**, *114*, 9987–10043.

3.　Schneider, J.; Matsuoka, M.; Takeuchi, M.; Zhang, J.; Horiuchi, Y.; Anpo, M.; Bahnemann, D.W. Understanding TiO$_2$ Photocatalysis: Mechanisms and Materials. *Chem. Rev.* **2014**, *114*, 9919–9986.

4.　Tong, H.; Ouyang, S.; Bi, Y.; Umezawa, N.; Oshikiri, M.; Ye, J. Nano-photocatalytic Materials: Possibilities and Challenges. *Adv. Mater.* **2012**, *24*, 229–251.

5. Cargnello, M.; Gordon, T.R.; Murray, C.B. Solution-Phase Synthesis of Titanium Dioxide Nanoparticles and Nanocrystals. *Chem. Rev.* **2014**, *114*, 9319–9345.

6. Tang, J.; Redl, F.; Zhu, Y.; Siegrist, T.; Brus, L.E.; Steigerwald, M.L. An Organometallic Synthesis of $TiO_2$ Nanoparticles. *Nano Lett.* **2005**, *3*, 543–548.

7. Chen, C.; Hu, R.; Mai, K.; Ren, Z.; Wang, H.; Qian, G.; Wang, Z. Shape Evolution of Highly Crystalline Anatase $TiO_2$ Nanobipyramids. *Cryst. Growth Des.* **2011**, *11*, 5221–5226.

8. Barnard, A.S.; Zapol, P. Effects of Particle Morphology and Surface Hydrogenation on the Phase Stability of $TiO_2$. *Phys. Rev. B* **2004**, *70*, 235403.

9. Rajh, T.; Dimitrijevic, N.M.; Bissonnette, M.; Koritarov, T.; Konda, V. Titanium Dioxide in the Service of the Biomedical Revolution. *Chem. Rev.* **2014**, *114*, 10177–10216.

10. De Angelis, F.; Di Valentin, C.; Fantacci, S.; Vittadini, A.; Selloni, A. Theoretical Studies on Anatase and Less Common $TiO_2$ Phases: Bulk, Surfaces, and Nanomaterials. *Chem. Rev.* **2014**, *114*, 9708–9753.

11. Hummer, D.R.; Kubicki, J.D.; Kent, P.R.C.; Post, J.E.; Heaney, P.J. Origin of Nanoscale Phase Stability Reversals in Titanium Oxide Polymorphs. *J. Phys. Chem. C* **2009**, *113*, 4240–4245.

12. Nunzi, F.; E. Mosconi, E.; Storchi, L.; Ronca, E.; Selloni, A.; Gratzel, M.; De Angelis, F. Inherent Electronic Trap States in $TiO_2$ Nanocrystals: Effect of saturation and sintering. *Energy Environ. Sci.* **2013**, *6*, 1221–1229.

13. Li, Y.-F.; Liu, Z.-P. Particle Size, Shape and Activity for Photocatalysis on Titania Anatase Nanoparticles in Aqueous Surroundings. *J. Am. Chem. Soc.* **2011**, *133*, 15743–15752.

14. Mattioli, G.; Bonapasta, A.A.; Bovi, D.; Giannozzi, P. Photocatalytic and Photovoltaic Properties of $TiO_2$ Nanoparticles Investigated by Ab Initio Simulations. *J. Phys. Chem. C* **2014**, *118*, 29928–29942.

15. Nunzi, F.; Storchi, L.; Manca, M.; Giannuzzi, R.; Gigli, G.; De Angelis, F. Shape and Morphology Effects on the Electronic Structure of $TiO_2$ Nanostructures: From Nanocrystals to Nanorods. *Appl. Mater. Interfaces* **2014**, *6*, 2471–2478.

16. Lamiel-Garcia, O.; Ko, K.C.; Lee, J.Y.; Bromley, S.T.; Illas, F. When Anatase Nanoparticles Become Bulklike: Properties of Realistic $TiO_2$ Nanoparticles in the 1–6 nm Size Range from All Electron Relativistic Density Functional Theory Based Calculations. *J. Chem. Theory Comput.* **2017**, *13*, 1785–1793.

17. Fazio, G.; Ferrighi, L.; Di Valentin, C. Spherical versus Faceted Anatase $TiO_2$ Nanoparticles: A Model Study of Structural and Electronic Properties. *J. Phys. Chem. C* **2015**, *119*, 20735–20746.

18. Selli, D.; Fazio, G.; Di Valentin, C. Modelling Realistic $TiO_2$ Nanospheres: A Benchmark Study of SCC-DFTB against Hybrid DFT. *J. Chem. Phys.* **2017**, *147*, 164701,.

19. Elstner, M.; Porezag, D.; Jungnickel, G.; Elsner, J.; Haugk, M.; Frauenheim, T.; Suhai, S.; Seifert, G. Self-Consistent-Charge Density-Functional Tight-Binding Method for Simulations of Complex Materials Properties. *Phys. Rev. B* **1998**, *58*, 7260.

20. Corà, F.; Alfredsson, M.; Mallia, G.; Middlemiss, D.S.; Mackrodt, W.C.; Dovesi, R.; Orlando, R. The Performance of Hybrid Density Functionals in Solid State Chemistry. In *Principles and Applications of Density Functional Theory in Inorganic Chemistry II. Structure and Bonding*; Springer: Berlin/Heidelberg, Germany, 2004; Volume 113.

21. Muscat, J.; Wander, A.; Harrison, N.M. On the Prediction of Band Gaps from Hybrid Functional Theory. *Chem. Phys. Lett.* **2001**, *3*, 397–401.

22. Labat, F.; Baranek, P.; Adamo, C. Structural and Electronic Properties of Selected Rutile and Anatase $TiO_2$ Surfaces: An ab Initio Investigation. *J. Chem. Theory. Comput.* **2008**, *4*, 341–352.

23. Dolgonos, G.; Aradi, B.; Moreira, N.H.; Frauenheim, T. An Improved Self-Consistent-Charge Density-Functional Tight-Binding (SCC-DFTB) Set of Parameters for Simulation of Bulk and Molecular Systems Involving Titanium. *J. Chem. Theory Comput.* **2010**, *6*, 266–278.

24. Luschtinetz, R.; Frenzel, J.; Milek, T.; Seifert, G. Adsorption of Phosphonic Acid at the $TiO_2$ Anatase (101) and Rutile (110) Surfaces. *J. Phys. Chem. C* **2009**, *113*, 5730–5740.

25. Fox, H.; Newman, K.E.; Schneider, W.F.; Corcelli, S.A. Bulk and Surface Properties of Rutile $TiO_2$ from Self-Consistent-Charge Density Functional Tight Binding. *J. Chem. Theory Comput.* **2010**, *6*, 499–507.

26. Fuertes, V.C.; Negre, C.F. A.; Oviedo, M.B.; Bonafé, F.P.; Oliva, F.Y.; Sánchez, C.G. A Theoretical Study of the Optical Properties of Nanostructured $TiO_2$. *J. Phys. Condens. Matter* **2013**, *25*, 115304.

27. Di Valentin, C.; Selloni, A. Bulk and Surface Polarons in Photoexcited Anatase $TiO_2$. *J. Phys. Chem. Lett.* **2011**, *2*, 2223–2228.

28. Fazio, G.; Ferrighi, L.; Di Valentin, C. Photoexcited Carriers Recombination and Trapping in Spherical vs Faceted TiO₂ Nanoparticles. *Nano Energy* **2016**, *27*, 673–689.

29. Nunzi, F.; Agrawal, S.; Selloni, A.; De Angelis, F. Structural and Electronic Properties of Photoexcited TiO₂ Nanoparticles from First Principles. *J. Chem. Theory Comput.* **2015**, *11*, 635–645.

30. Nunzi, F.; De Angelis, F.; Selloni, A. Ab Initio Simulation of the Absorption Spectra of Photoexcited Carriers in TiO₂ Nanoparticles. *J. Phys. Chem. Lett.* **2016**, *7*, 3597–3602.

31. Diebold, U. The Surface Science of Titanium Dioxide. *Surf. Sci. Rep.* **2003**, *48*, 53–229.

32. Sclafani, A.; Herrmann, J.M. Comparison of the Photoelectronic and Photocatalytic Activities of Various Anatase and Rutile Forms of Titania in Pure Liquid Organic Phases and in Aqueous Solutions. *J. Phys. Chem.* **1996**, *100*, 13655–13661.

33. Dimitrijevic, N.M.; Vijayan, B.K.; Poluektov, O.G.; Rajh, T.; Gray, K.A.; He, H.; Zapol, P. Role of Water and Carbonates in Photocatalytic Transformation of CO₂ to CH4 on Titania. *J. Am. Chem. Soc.* **2011**, *133*, 3964–3971.

34. Mino, L.; Zecchina, A.; Martra, G.; Rossi, A.M.; Spoto, G. A Surface Science Approach to TiO₂ P25 Photocatalysis: An in Situ FTIR Study of Phenol Photodegradation at Controlled Water Coverages from Sub-Monolayer to Multilayer. *Appl. Catal. B Environ.* **2016**, *196*, 135–141.

35. Shirai, K.; Sugimoto, T.; Watanabe, K.; Haruta, M.; Kurata, H.; Matsumoto, Y. Effect of Water Adsorption on Carrier Trapping Dynamics at the Surface of Anatase TiO₂ Nanoparticles. *Nano Lett.* **2016**, *16*, 1323–1327.

36. Panarelli, E.G.; Livraghi, S.; Maurelli, S.; Polliotto, V.; Chiesa, M.; Giamello, E. Role of Surface Water Molecules in Stabilizing Trapped Hole Centres in Titanium Dioxide (Anatase) as Monitored by Electron Paramagnetic Resonance. *J. Photochem. Photobiol. A* **2016**, *322*, 27–34.

37. Addamo, M.; Augugliaro, V.; Coluccia, S.; Di Paola, A.; García-López, E.; Loddo, V.; Marcì, G.; Martra, G.; Palmisano, L. The Role of Water in the Photocatalytic Degradation of Acetonitrile and Toluene in Gas-Solid and Liquid-Solid Regimes. *Int. J. Photoenergy* **2006**, *2006*, 39182.

38. Miller, K.L.; Lee, C.W.; Falconer, J.L.; Medlin, J.W. Effect of Water on Formic Acid Photocatalytic Decomposition on TiO₂ and Pt/TiO₂. *J. Catal.* **2010**, *275*, 294–299.

39. Salvador, P. On the Nature of Photogenerated Radical Species Active in the Oxidative Degradation of Dissolved Pollutants with TiO₂ Aqueous Suspensions: A Revision in the Light of the Electronic Structure of Adsorbed Water. *J. Phys. Chem. C* **2007**, *111*, 17038–17043.

40. Selli, D.; Fazio, G.; Seifert, G.; Di Valentin, C. Water Multilayers on TiO₂ (101) Anatase Surface: Assessment of a DFTB-Based Method. *J. Chem. Theory Comput.* **2017**, *13*, 3862–3873.

41. Dovesi, R.; Saunders, V.R.; Roetti, C.; Orlando, R.; Zicovich-Wilson, C.M.; Pascale, F.; Civalleri, B.; Doll, K.; Harrison, N.M.; Bush, I.J.; et al. *CRYSTAL14 User's Manual*; University of Torino: Torino, Italy, 2014.

42. Becke, A.D. Density-Functional Thermochemistry. III. The Role of Exact Exchange. *J. Chem. Phys.* **1993**, *98*, 5648.

43. Lee, C.; Yang, W.; Parr, R.G. Development of the Colle-Salvetti Correlation-Energy Formula into a Functional of the Electron Density. *Phys. Rev. B* **1988**, *37*, 785–789.

44. Krukau, A.V.; Vydrov, O.A.; Izmaylov, A.F.; Scuseria, G.E. Influence of the Exchange Screening Parameter on the Performance of Screened Hybrid Functionals. *J. Chem. Phys.* **2006**, *125*, 224106.

45. Giannozzi, P.; Baroni, S.; Bonini, N.; Calandra, M.; Car, R.; Cavazzoni, C.; Ceresoli, D.; Chiarotti, G.L.; Cococcioni, M.; Dabo, I.; et al. QUANTUM ESPRESSO: A Modular and Open-Source Software Project for Quantum Simulations of Materials. *J. Phys. Condens. Matter* **2009**, *21*, 395502.

46. Perdew, J.P.; Burke, K.; Ernzerhof, M. Generalized Gradient Approximation Made Simple. *Phys. Rev. Lett.* **1996**, *77*, 3865–3868.

47. Burdett, J.K.; Hughbanks, T.; Miller, G.J.; Richardson, J.W., Jr.; Smith, J.V. Structural-Electronic Relationships in Inorganic Solids: Powder Neutron Diffraction Studies of the Rutile and Anatase Polymorphs of Titanium Dioxide at 15 K and 295 K. *J. Am. Chem. Soc.* **1987**, *109*, 3639–3646.

48. Elstner, M.; Seifert, G. Density Functional Tight Binding. *Philos. Trans. R. Soc. A* **2014**, *372*, 20120483.

49. Seifert, G.; Joswig, J.-O. Density-Functional Tight Binding–an Approximate Density-Functional Theory Method. *WIREs Comput. Mol. Sci.* **2012**, *2*, 456–465.

50. Aradi, B.; Hourahine, B.; Frauenheim, T. DFTB+, a Sparse Matrix-Based Implementation of the DFTB Method. *J. Phys. Chem. A* **2007**, *111*, 5678–5684.

51. Hu, H.; Lu, Z.; Elstner, M.; Hermans, J.; Yang, W. Simulating Water with the Self-Consistent-Charge Density Functional Tight Binding Method: From Molecular Clusters to the Liquid State. *J. Phys. Chem. A* **2007**, *111*, 5685.

52. Tang, H.; Levy, F.; Berger, H.; Schmid, P.E. Urbach tail of anatase TiO₂. *Phys. Rev. B* **1995**, *52*, 7771.

53. Liu, Y.; Claus, R.O. Blue Light Emitting Nanosized TiO₂ Colloids. *J. Am. Chem. Soc.* **1997**, *119*, 5273–5274.

54. Koch, S.W.; Kira, M.; Khitrova, G.; Gibbs, H.M. Semiconductor Excitons in New Light. *Nat. Mater.* **2006**, *5*, 523–531.

55. Tang, H.; Berger, H.; Schmid, P.E.; Lévy, F. Photoluminescence in TiO₂ Anatase Single Crystals. *Solid State Commun.* **1993**, *87*, 847–850.

56. El-Sayed, M.A. Small Is Different: Shape-, Size-, and Composition-Dependent Properties of Some Colloidal Semiconductor Nanocrystals. *Acc. Chem. Res.* **2004**, *37*, 326–333.

57. Mori-Sánchez, P.; Cohen, A.J.; Yang, W. Localization and Delocalization Errors in Density Functional Theory and Implications for Band-Gap Prediction. *Phys. Rev. Lett.* **2008**, *100*, 146401

58. Deskins, N.A.; Dupuis, M. Intrinsic Hole Migration Rates in TiO₂ from Density Functional Theory. *J. Phys. Chem. C* **2009**, *113*, 346–358.

59. Najafov, H.; Tokita, S.; Ohshio, S.; Kato, A.; Saitoh, H. Green and Ultraviolet Emissions from Anatase TiO₂ Films Fabricated by Chemical Vapor Deposition. *Jpn. J. Appl. Phys.* **2005**, *44*, 245–253.

60. Panayotov, D.A., Yates Jr, J.T. n-Type Doping of TiO₂ with Atomic Hydrogen - Observation of the Production of Conduction Band Electrons by Infrared Spectroscopy. *Chem. Phys. Lett.* **2007**, *436*, 204.

61. Yamakata, A.; Ishibashi, T.; Onishi, H. Time-Resolved Infrared Absorption Spectroscopy of Photogenerated Electrons in Platinized TiO₂ Particles. *Chem. Phys. Lett.* **2001**, *333*, 271–277.

62. Durrant, J.R. Modulating Interfacial Electron Transfer Dynamics in Dye Sensitised Nanocrystalline Metal Oxide Films. *J. Photochem. Photobiol. A* **2002**, *148*, 5–10.

63. Martin, S.T.; Hermann, H.; Hoffmann, M.R. Time-Resolved Microwave Conductivity. Part 1.—TiO₂ Photoreactivity and Size Quantization. *J. Chem. Soc. Faraday Trans.* **1994**, *90*, 3323–3330.

64. Beermann, N.; Boschloo, G.; Hagfeldt, A. Trapping of electrons in nanostructured TiO₂ studied by photocurrent transients. *J. Photochem. Photobiol. A* **2002**, *152*, 213.

65. Boschloo, G.; Fitzmaurice, D. Spectroelectrochemical Investigation of Surface States in Nanostructured TiO₂ Electrodes. *J. Phys. Chem. B* **1999**, *103*, 2228–2231.

66. Szezepankiewicz, S.H.; Moss, J.A.; Hoffmann, M.R. Slow Surface Charge Trapping Kinetics on Irradiated TiO₂. *J. Phys. Chem. B* **2002**, *106*, 2922–2927.

67. Maurelli, S.; Livraghi, S.; Chiesa, M.; Giamello, E.; Van Doorslaer, S.; Di Valentin, C.; Pacchioni, G. Hydration Structure of the Ti(III) Cation as Revealed by Pulse EPR and DFT Studies: New Insights into a Textbook Case. *Inorg. Chem.* **2011**, *50*, 2385–2394.

68. Chiesa, M.; Paganini, M.C.; Livraghi, S.; Giamello, E. Charge Trapping in TiO₂ Polymorphs as Seen by Electron Paramagnetic Resonance Spectroscopy. *Phys. Chem. Chem. Phys.* **2013**, *15*, 9435–9447.

69. Micic, O.I.; Zhang, Y.; Cromack, K.R.; Trifunac, A.D.; Thurnauer, M.C. Trapped Holes on Titania Colloids Studied by Electron Paramagnetic Resonance. *J. Phys. Chem.* **1993**, *97*, 7277–7283.

70. Brezová, V.; Barbieriková, Z.; Zukalová, M.; Dvoranová, D.; Kavan, L. EPR Study of 17O-Enriched Titania Nanopowders under UV Irradiation. *Catal. Today* **2014**, *230*, 112–118.

71. Gallino, F.; Pacchioni, G.; Di Valentin, C. Transition Levels of Defect Centers in ZnO by Hybrid Functionals and Localized Basis Set Approach. *J. Chem. Phys.* **2010**, *133*, 144512.

72. Shkrob, I.A.; Sauer, M.C., Jr. Hole Scavenging and Photo-Stimulated Recombination of Electron−Hole Pairs in Aqueous TiO₂ Nanoparticles. *J. Phys. Chem. B* **2004**, *108*, 12497–12511.

73. He, Y.; Tilocca, A.; Dulub, O.; Selloni, A.; Diebold, U. Local Ordering and Electronic Signatures of Submonolayer Water on Anatase TiO₂(101). *Nat. Mater.* **2009**, *8*, 585–589.

74. Tilocca, A.; Selloni, A. Vertical and Lateral Order in Adsorbed Water Layers on Anatase TiO₂(101). *Langmuir* **2004**, *20*, 8379–8384.

75. Tilocca, A.; Selloni, A. DFT-GGA and DFT+U Simulations of Thin Water Layers on Reduced TiO₂ Anatase. *J. Phys. Chem. C* **2012**, *116*, 9114–9121.

76. Aschauer, U.J.; Tilocca, A.; Selloni, A. Ab Initio Simulations of the Structure of Thin Water Layers on Defective Anatase TiO₂ (101) Surfaces. *Int. J. Quant. Chem.* **2015**, *115*, 1250–1257.

77. Zhu, Y.; Ding, C.; Ma, G.; Du, Z. Electronic State Characterization of TiO₂ Ultrafine Particles by Luminescence Spectroscopy. *J. Solid State Chem.* **1998**, *139*, 124–127.

78. Dimitrijevic, N.M.; Saponjic, Z.V.; Rabatic, B.M.; Poluektov, O.G.; Rajh T. Effect of Size and Shape of Nanocrystalline TiO$_2$ on Photogenerated Charges. An EPR Study. *J. Phys. Chem. C* **2007**, *111*, 14597–14601.

79. Luca, V. Comparison of Size-Dependent Structural and Electronic Properties of Anatase and Rutile Nanoparticles. *J. Phys. Chem. C* **2009**, *113*, 6367–6380.

80. Colombo Jr., D.P.; Roussel, K.A.; Saeh, J.; Skinner, D.E.; Cavaleri, J.J.; Bowman R.M. Femtosecond Study of the Intensity Dependence of Electron-Hole Dynamics in TiO$_2$ Nanoclusters. *Chem. Phys. Lett.* **1995**, *232*, 207–214.

81. Lazzeri, M.; Vittadini, A.; Selloni, A. Structure and Energetics of Stoichiometric TiO$_2$ Anatase Surfaces. *Phys. Rev. B* **2001**, *63*, 155409.

82. Labat, F.; Baranek, P.; Domain, C.; Minot, C.; Adamo, C. Density Functional Theory Analysis of the Structural and Electronic Properties of TiO$_2$ Rutile and Anatase Polytypes: Performances of Different Exchange-Correlation Functionals. *J. Chem. Phys.* **2007**, *126*, 154703.

83. Xu, X.; Goddard, W.A. Bonding Properties of the Water Dimer: A Comparative Study of Density Functional Theories. *J. Phys. Chem. A* **2004**, *108*, 2305–2313.

84. Tschumper, G.S.; Leininger, M.L.; Hoffman, B.C.; Waleev, E.F.; Schaefer, H.F., III; Quack, M. Anchoring the Water Dimer Potential Energy Surface with Explicitly Correlated Computations and Focal Point Analyses. *J. Chem. Phys.* **2002**, *116*, 690–701.

85. Curtiss, L.A.; Frurip, D.J.; Blander, M. Studies of Molecular Association in H$_2$O and D$_2$O Vapors by Measurement of Thermal Conductivity. *J. Chem. Phys.* **1979**, *71*, 2703–2711.

86. Odutola, J.A.; Dyke, T.R. Partially Deuterated Water Dimers: Microwave Spectra and Structure. *J. Chem. Phys.* **1980**, *72*, 5062–5070.

87. Soper, A.K.; Benmore, C.J. Quantum Differences between Heavy and Light Water. *Phys. Rev. Lett.* **2008**, *101*, 065502.

88. Vittadini, A.; Selloni, A.; Rotzinger, F.P.; Grätzel, M. Structure and Energetics of Water Adsorbed at TiO$_2$ Anatase (101) and (001) Surfaces. *Phys. Rev. Lett.* **1998**, *81*, 2954–2957.

89. Egashira, M.; Kawasumi, S.; Kagawa, S.; Seiyama, T. Temperature Programmed Desorption Study of Water Adsorbed on Metal Oxides. I. Anatase and Rutile. *Bull. Chem. Soc. Jpn.* **1978**, *51*, 3144–3149.

90. Beck, D.D.; White, J.M.; Ratcliffe, C.T. Catalytic Reduction of CO with Hydrogen Sulfide. 2. Adsorption of H$_2$O and H$_2$S on Anatase and Rutile. *J. Phys. Chem.* **1986**, *90*, 3123–3131.

91. Herman, G.S.; Dohnàlek, Z.; Ruzycki, N.; Diebold, U. Experimental Investigation of the Interaction of Water and Methanol with Anatase-TiO$_2$(101). *J. Phys. Chem. B* **2003**, *107*, 2788–2795.

92. Zhao, Z.; Li, Z.; Zou, Z. Structure and Properties of Water on the Anatase TiO$_2$(101) Surface: From Single-Molecule Adsorption to Interface Formation. *J. Phys. Chem. C* **2012**, *116*, 11054–11061.

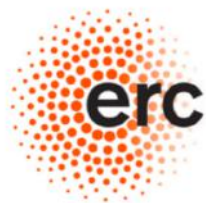

European Research Council
Established by the European Commission

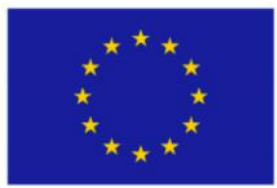

# Using Density Functional Theory to Model Realistic TiO₂ Nanoparticles, Their Photoactivation and Interaction with Water.

Daniele Selli, Gianluca Fazio and Cristiana Di Valentin[*]

Dipartimento di Scienza dei Materiali, Università di Milano-Bicocca,

via R.Cozzi 55 20125, Milano, Italy

**SUPPLEMENTARY MATERIAL**

---

[*] E-mail address: cristiana.divalentin@unimib.it.
Phone: +390264485235. Fax: +390264485400

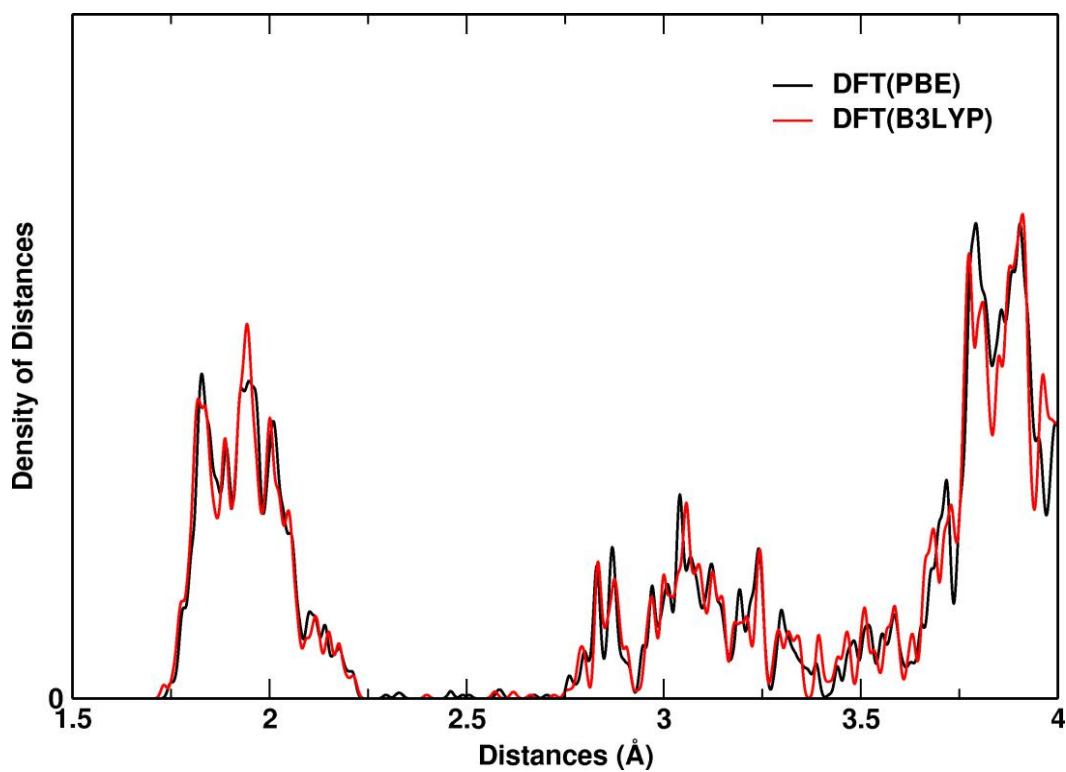

**Figure S1**. Comparison of the distances distribution (simulated EXAFS) computed with DFT(PBE), in black and DFT(B3LYP) in red, for the 2.2 nm NS produced at 300 K.

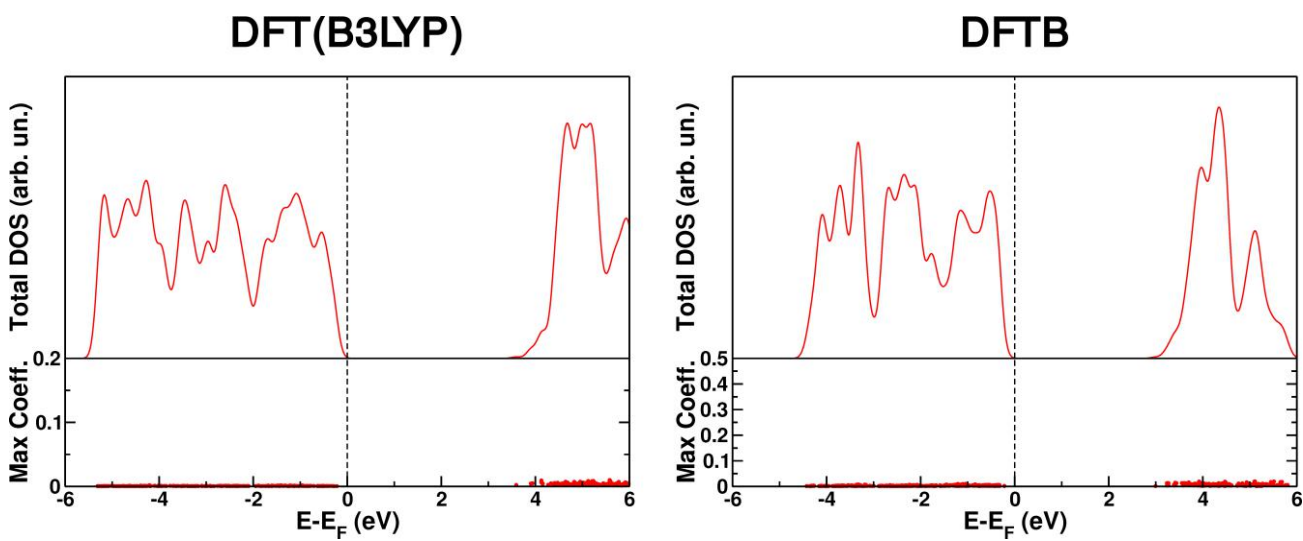

**Figure S2**. DFT(B3LYP) and DFTB total (DOS) density of states for anatase bulk $TiO_2$. The maximum atomic orbital coefficient ($max_c$) of each eigenstate is also reported.